\documentclass[11pt]{article}

\newcommand{\Hp}{$H^+$}
\newcommand{\Gp}{$G^+$}

\newcommand{\bfr}{\begin{flushright}}
\newcommand{\efr}{\end{flushright}}
\newcommand{\bfl}{\begin{flushleft}}
\newcommand{\efl}{\end{flushleft}}

\title{\bf On the Renormalization of \\
Two-Higgs-Doublet Models \\ [0.25cm]}
\author{R. Santos${}^{1,2}$ and A. Barroso \\[0.25cm]
{\normalsize Dept. de F\'\i sica}\\
{\normalsize Faculdade de Ci\^encias}\\
{\normalsize Universidade de Lisboa}\\
{\normalsize Campo Grande, C1, 1700 Lisboa}\\
{\normalsize Portugal}\\[0.5cm]}

\date{December 1996 \\[0.1cm]}

\begin{document}

\maketitle

\begin{abstract}

In this paper we perform the complete one-loop renormalization of a general Two-Higgs-Doublet Model. We present all the vertices for this model including the ones in the scalar sector and calculate all the counterterms of the theory.

\end{abstract}

\footnotetext[1]{e-mail:fsantos@skull.cc.fc.ul.pt}
\footnotetext[2]{Partially supported by JNICT contract BD/2077/92-RM}

\thispagestyle{empty}

\newpage

\section{Introduction}  

When the LEP accelerator at CERN enters the second phase of its program, the SU(2)$\otimes$U (1) standard model does not need any more praise. The theory has been successfully scrutinized and the agreement between its predictions and the experimental results is impressive (e.g. ref. \cite{SMP}). Besides the effort of large teams of devoted experimenters, this endeavour also required a number of detailed calculations beyond the lowest order of perturbation theory. Hence, one can say that the renormalization of the SU(2)$\otimes$U(1) theory has passed from the formal stage of its establishment \cite{Ren} into the world of practical calculations. For this purpose it is very useful to have the review article of Aoki et al. \cite{Aoki} which can be considered as a good SU(2)$\otimes$U(1) practitioner guide. So far, it seems that such a guide does not exist for the two-Higgs-doublet models (2HDM). This is the aim of this article.

Several reasons can be given to justify the study of the standard model with two doublets. In our opinion, the best reason is the fact that there is no information about the Higgs sector. Hence, given the crucial role that the scalar sector plays in the theory, it is at least prudent to explore reasonable extensions of the minimal Higgs sector.

Over the last few years, a great deal of work has been invested in the study of several production and decay mechanisms associated with the Higgs bosons of the 2HDM. Fortunately, this large amount of work is beautifully and systematically presented in the Higgs Hunter's Guide \cite{GHDK90}, which we shall consider as our basic reference for the work done until the end of 1989.

Several authors have performed one-loop calculations in the 2HDM. \mbox{After} the ex\-pe\-rimental evidence for a top quark mass \cite{CDF}, M\'endez and Po\-ma\-rol \cite{MP92} have computed, in the unitary gauge, the  \(O({m_t}^2/{M_W}^2) \) corrections to the hadronic width of the Higgs bosons. In the minimal super\-symmetric standard model (MSSM) several authors \cite{RR92} have estimated the process \(H^+ \rightarrow W^+ \gamma\) which is forbidden to occur at tree level. Because of this fact the calculation can be done, including all reducible and irreducible 3-point functions and do not require the specification of the renormalization scheme and the calculation of the counterterms. Another relevant work with a great deal of details about the renormalization of the MSSM is the article by Pierce and Papadoupolos \cite{PP93} where they have considered one loop cor\-rec\-tions to the decay  \(H \rightarrow ZZ\). However, to preserve the mass sum rule for the renor\-ma\-lized masses of the neutral Higgs bosons, they introduce a \(\overline{MS}\) scheme to renormalize the angle $\beta$. Clearly, this is not entirely consistent with the on-shell scheme and furthermore it is not valid in the general 2HDM. A systematic on-shell renormalization study for the Higgs and gauge boson sectors of the MSSM was carried out by Chankowski, Pokorski and Rosiek \cite{CPR94}. Here we present a similar work for a general 2HDM. The potential depends on seven real parameters rather than three as is the case for the MSSM. On the other hand, instead of renormalizing the pa\-ram\-e\-ters of the potential, as was done by Chankowski et al. \cite{CPR94}, we renormalize the masses $m_H$, $m_h$, $m_A$, and ${m_H}^+$ and the angles $\beta=tan(v_2/v_1)$ and  $\alpha$.

\section{The Higgs potential}

To define our notation we start with a brief review of the two-Higgs-doublet potential. Let $\phi_i$, with i=1,2, denote two complex scalar doublets with hyper\-charge Y=1. Introducing the complete set of invariants $x_1=\phi_1^{\dagger} \phi_1$,  $x_2=\phi_2^{\dagger} \phi_2$, $x_3=Re \{ \phi_1^{\dagger} \phi_2 \}$ and $x_4=Im \{ \phi_1^{\dagger} \phi_2 \}$, it is clear that the most general SU(2) $\otimes$ U(1) invariant renormalizable potential depends on 14 real parameters and can be written in the form

\begin{equation}
V=-\sum_{i=1}^{4} \mu_{i}^2 x_{i} + \sum_{i \leq j = 1}^{4} b_{ij} x_{i} x_{j}\enskip .
\end{equation}
Under CP the fields transform as

\begin{equation}
\phi_{i} \rightarrow e^{i \alpha_{i}} \phi_{i}^*
\end{equation}
with arbitrary phase \(\alpha_{i}\). Choosing these phases to be zero, it is immediate to conclude that an explicit CP conserving potential, $V_{CP}$, has $\mu_{4}^2=b_{14}=b_{24}=$ $b_{34}=0$. Hence,  $V_{CP}$ depends on 10 real arbitrary parameters. How\-ever, such a potential could still break CP spontaneously \cite{L73}. In a previous paper \cite{rui1} we have shown that there are two possibilities to impose in a natural way that the potential has only CP invariant minima. These require \mbox{$b_{13}=b_{23}=0$} and either $\mu_{3}^2=0$ and $b_{33}\neq b_{44}$ or $\mu_{3}^2 \neq 0$ and $b_{33}=b_{44}$. Here we shall use the first version of the potential which we rewrite in the form:

\begin{equation}
V=-\mu_{1}^{2} x_{1}-\mu_{2}^{2} x_{2}+\lambda_{1} x_{1}^2+\lambda_{2} x_{2}^2+\lambda_{3} x_{3}^2+\lambda_{4} x_{4}^2+\lambda_{5} x_{1} x_{2} \enskip .
\end{equation}
Notice that this 7-parameter potential obeys the discrete symmetry \mbox{\(\phi_1 \rightarrow -\phi_1\)} which is usually introduced to guarantee the absence of flavour-changing neutral currents (FCNC) in the tree-level Yukawa couplings. It is in\-ter\-esting to point out \cite{rui1} that potentials with only CP invariant minima are consistent with the absence of FCNC in the fermionic sector.
Now, denoting by  \(v_i/ \sqrt{2}\) the vacuum expectation value of each of the two doublets, we can write  \(\phi_i\) in the form

\begin{equation}
\phi_{i}=
\left[
 \begin{array}{c}
a_{i}^+  \\ (v_{i}+b_{i}+ic_{i})/ \sqrt{2} 
 \end{array} \right]
\end{equation}
where \(a_{i}^+\) are complex fields, and \(b_i\) and \(c_i\) are real fields. This, in turn, enables us to rewrite the potential (3) as:

\begin{eqnarray}
V & = & -\frac{\rm 1}{\rm 2} \lambda_3
\left[
\begin{array}{c}
a_{1}^+ \quad a_{2}^+ 
\end{array} \right]
M_{\beta}
\left[
\begin{array}{c}
a_{1}^- \\ a_{2}^- 
\end{array} \right] \nonumber \\ & & \nonumber \\
& & +\frac{1}{4} (\lambda_4-\lambda_3)
\left[
\begin{array}{c}
c_{1} \quad c_{2} 
\end{array} \right]
M_{\beta}
\left[
\begin{array}{c}
c_{1} \\ c_{2} 
\end{array} \right]
+\frac{\rm 1}{\rm 2}
\left[
\begin{array}{c}
b_{1}\quad b_{2} 
\end{array} \right]
M_{\alpha}
\left[
\begin{array}{c}
b_{1} \\ b_{2} 
\end{array} \right] \nonumber \\ & & \nonumber \\
& & + \left[
\begin{array}{c}
a_{1}^+ \quad a_{2}^+ 
\end{array} \right]
T
\left[
\begin{array}{c}
a_{1}^- \\ a_{2}^- 
\end{array} \right]
+\frac{\rm 1}{\rm 2}
\left[
\begin{array}{c}
c_{1}\quad c_{2} 
\end{array} \right]
T
\left[
\begin{array}{c}
c_{1} \\ c_{2} 
\end{array} \right] \\ & & \nonumber \\ 
& &+\frac{\rm 1}{\rm 2}
\left[
\begin{array}{c}
b_{1}\quad b_{2} 
\end{array} \right]
T
\left[
\begin{array}{c}
b_{1} \\ b_{2} 
\end{array} \right] \nonumber \\ & & \nonumber \\
& & + T_1 b_1 + T_2 b_2 + \enskip cubic \quad and \quad quartic \quad terms \nonumber
\end{eqnarray}
with the matrices $M_{\beta}$, $M_{\alpha}$ and T defined as

\renewcommand{\theequation}{\arabic{equation}\mbox{a}}
\begin{equation}
M_{\beta}=
\left[
\begin{array}{cc}
v_2^2 &  -v_1 v_2 \\
- v_1 v_2 &  v_1^2
\end{array} \right]
\end{equation}
\addtocounter{equation}{-1}
\renewcommand{\theequation}{\arabic{equation}\mbox{b}}
\begin{equation}
M_{\alpha}=
\left[
\begin{array}{cc}
2v_1^2 \lambda_1 &  v_1 v_2 (\lambda_3  + \lambda_5) \\
v_1 v_2 (\lambda_3 + \lambda_5) & 2 v_2^2 \lambda_2
\end{array} \right]
\end{equation}
\addtocounter{equation}{-1}
\renewcommand{\theequation}{\arabic{equation}\mbox{c}}
\begin{equation}
T=
\left[
\begin{array}{cc}
\frac{T_1}{v_1} & 0 \\
0 &  \frac{T_2}{v_2}
\end{array} \right]
\end{equation}
with
\renewcommand{\theequation}{\arabic{equation}\mbox{a}}
\begin{equation}
T_1 \, = \, v_1(-\mu_1^2 +\lambda_1 v_1^2+\frac{\lambda_3+\lambda_5}{2}v_2^2)
\end{equation}
\addtocounter{equation}{-1}
\renewcommand{\theequation}{\arabic{equation}\mbox{b}}
\begin{equation}
T_2 \, = \,v_2(-\mu_2^2 +\lambda_2 v_2^2+\frac{\lambda_3+\lambda_5}{2}v_1^2) \enskip .
\end{equation}

The conditions for a local extreme of the potential are $T_1=T_2=0$. Diagonalizing the quadratic terms of V one obtains the mass eigenstates: 2 neutral CP-even scalar particles, H and h, a neutral CP-odd scalar particle, A, and the would-be Goldstone boson partner of the Z, $G_0$, a charged Higgs field \Hp and the Goldstone associated with the W boson, \Gp. The relations between the mass eigenstates and the SU(2)$\otimes$U(1) eigenstates are: 

\renewcommand{\theequation}{\arabic{equation}\mbox{a}}
\begin{equation}
\left[
\begin{array}{c}
H \\ h
\end{array} \right]
=R_{\alpha}
\left[
\begin{array}{c}
b_1 \\ b_2
\end{array} \right]
\end{equation}
\addtocounter{equation}{-1} \\
\renewcommand{\theequation}{\arabic{equation}\mbox{b}}
\begin{equation}
\left[
\begin{array}{c}
H^+ \\ G^+
\end{array} \right]
=R_{\beta}
\left[
\begin{array}{c}
a_1^+ \\ a_2^+
\end{array} \right]
\end{equation}
\addtocounter{equation}{-1} \\
\renewcommand{\theequation}{\arabic{equation}\mbox{c}}
\begin{equation}
\left[
\begin{array}{c}
A \\ G_0
\end{array} \right]
=R_{\beta}
\left[
\begin{array}{c}
c_1 \\ c_2
\end{array} \right]
\end{equation}
with
\renewcommand{\theequation}{\arabic{equation} \mbox{a}}
\begin{equation}
R_{\alpha}=
\left[
\begin{array}{cc}
\cos \alpha & \sin \alpha \\
- \sin \alpha & \cos \alpha
\end{array} \right]
\end{equation}
\addtocounter{equation}{-1} \\
\renewcommand{\theequation}{\arabic{equation}\mbox{b}}
\begin{equation}
R_{\beta}=
\left[
\begin{array}{cc}
- \sin \beta & \cos \beta \\
\cos \beta & \sin \beta
\end{array} \right]
\end{equation}
\addtocounter{equation}{-1}
\renewcommand{\theequation}{\arabic{equation}\mbox{c}}
\begin{equation}
\tan \beta=\frac{v_2}{v_1} \qquad ; \qquad
\tan 2 \alpha=\frac{v_1v_2(\lambda_3+\lambda_5)}{\lambda_2 v_2^2-\lambda_1 v_1^2} \enskip .
\end{equation}

For the renormalization program it is convenient to rewrite V in terms of the mass eigenstates. After some straightforward algebra one obtains:

\renewcommand{\theequation}{\arabic{equation}}
\begin{eqnarray}
{\cal L} & = & -T_H H -T_h h - H^2
\left\{
\frac{M_H^2}{2}+\frac{T_{\alpha \beta}+T_{\delta}\sin^2 \alpha}{v \sin 2 \beta}
\right\} - h^2
\left\{
\frac{M_h^2}{2}+\frac{T_{\alpha \beta}+T_{\delta} \cos^2 \alpha}{v \sin 2 \beta}
\right\} \nonumber \\ & &  \nonumber \\
& & - H h
\left\{
\frac{T_{\delta} \sin 2 \alpha}{v \sin 2 \beta}
\right\} - A^2
\left\{
\frac{M_A^2}{2}+\frac{T_{\alpha \beta}+T_{\delta}\cos^2 \beta}{v \sin 2 \beta}
\right\} - G_0^2
\left\{
\frac{T_{\alpha \beta}+T_{\delta} \sin^2 \beta}{v \sin 2 \beta}
\right\}  \\ & & \nonumber \\
& & -A G_0
\left\{
\frac{T_{\delta}}{v}
\right\} -H^+ H^-
\left\{
M_{H^+}^2+2 \frac{T_{\alpha \beta}+T_{\delta}\cos^2 \beta}{v \sin 2 \beta}
\right\}  
- (H^+G^-+G^+H^-)
\left\{
\frac{T_{\delta}}{v}
\right\} \nonumber \\
& &  -G^+ G^-
\left\{
2 \frac{T_{\alpha \beta}+T_{\delta} \sin^2 \beta}{v \sin 2 \beta}
\right\} + \enskip cubic \quad and \quad quartic \quad terms \nonumber
\end{eqnarray}
with

\renewcommand{\theequation}{\arabic{equation}\mbox{a}}
\begin{equation}
\left[
\begin{array}{c}
T_H \\ T_h
\end{array} \right]
=R_{\alpha}
\left[
\begin{array}{c}
T_1 \\ T_2
\end{array} \right]
\end{equation}
\addtocounter{equation}{-1}
\renewcommand{\theequation}{\arabic{equation}\mbox{b}}
\begin{equation}
T_{\delta}=T_H \sin \delta + T_h \cos \delta
\end{equation}
\addtocounter{equation}{-1}
\renewcommand{\theequation}{\arabic{equation}\mbox{c}}
\begin{equation}
T_{\alpha \beta}=\sin \beta (T_H \cos \alpha - T_h \sin \alpha)
\end{equation}
and \(\delta=\alpha - \beta\).
As we have already pointed out, at tree-level, all T-terms are zero. So, at tree-level, the linear terms and the mixed terms vanish and the coefficients of the terms with quadratic fields, are, as they should be, their mass squared. However, at one-loop order these statements are no longer true, and this particular form of writing V will be useful in the derivation of the counterterms to renormalize some scalar particles Green's functions.

\section{The lagrangean}
\subsection{The classical lagrangean}

For completeness let us write the classical lagrangean of the standard model in the form:

\renewcommand{\theequation}{\arabic{equation}}
\begin{equation}
\cal {L_C}=\cal {L_{YM}} + \cal {L_F} + \cal {L_S} + \cal {L_Y}
\end{equation}
where \(\cal {L_{YM}}\) is the gauge boson sector of the model,  \(\cal {L_{F}}\) denotes the fermionic kinetic term and their couplings to the gauge bosons,  \(\cal {L_{S}}\) stands for the scalar sector of the theory and  \(\cal {L_{Y}}\) denotes the Yukawa couplings of fermion and scalar particles. The first two terms of eq. (12) are the same for the standard model and for the 2HDM and so there is no need to write them explicitly here. The scalar lagrangean is given by:

\begin{equation}
{\cal L_{S}} = \sum_{i=1}^{2}
\left(
D_{\mu}\phi_{i}
\right)^\dagger
D^\mu \phi_{i}-V
\left(
\phi_1, \phi_2
\right)
\end{equation}
where

\begin{equation}
D_{\mu}=\partial_{\mu}-i g_1 I^a W_{\mu}^a + i g_2 \frac{Y}{2}B_{\mu}
\end{equation}
is the covariant derivative and \(V \left(\phi_1, \phi_2 \right)\) is the potential that we have dis\-cussed in the previous paragraph. The Yukawa lagrangean is, again, a straightforward generalisation of the similar form in the standard model. In principle we could write all terms in \(\cal {L_{Y}}\) in the form

\begin{equation}
g_{ij}^k \left[u \quad d \right]_{L}^i \phi^k d_{R}^j
\end{equation}
where the \(g_{ij}^k\) are arbitrary Yukawa constants and i and j are quark gen\-er\-a\-tion indices. However, to avoid the existence of tree-level FCNC, one should impose the condition that the same scalar doublet  \(\phi^k\) does not couple to both up and down quarks. There are essentially four ways of doing this and so there are four variations of the model. A further discussion of this point, which is not relevant for the renormalization discussion, can be found in the Higgs Hunter's Guide \cite{GHDK90}. The four different models will be presented in Appendix A.

\subsection{The gauge fixing and ghost lagrangeans}

At the quantum level the action involves another contribution to the la\-grangean called the gauge fixing term,  \(\cal {L_{GF}}\). The existence of such a term is by now a textbook subject. So, we can simply state that calculations are easily done in the so-called linear  \(R_{\xi}\) gauges given by          

\begin{equation}
{\cal L_{GF}}  =  -\frac{1}{2 \xi_A}(\partial . A)^2-\frac{1}{2 \xi_Z}(\partial . Z- \xi_Z M_Z G_0)^2 -\frac{1}{ \xi_W}|\partial .W^+ +i \xi_W M_W G^+|^2
\end{equation}
where \(\xi_W\),  \(\xi_A\),  \(\xi_Z\) are arbitrary parameters and the Z and the photon field, A, are expressed in terms of the original gauge fields by the equations:

\renewcommand{\theequation}{\arabic{equation}\mbox{a}}
\begin{equation}
Z_{\mu} = \cos \theta_W W_{\mu}^3 + \sin \theta_W B_{\mu}
\end{equation}
\addtocounter{equation}{-1}
\renewcommand{\theequation}{\arabic{equation}\mbox{b}}
\begin{equation}
A_{\mu} = -\sin \theta_W W_{\mu}^3 + \cos \theta_W B_{\mu} \enskip .
\end{equation}
\\
Just for completeness let us recall that

\renewcommand{\theequation}{\arabic{equation}\mbox{a}}
\begin{equation}
M_{W} = \frac{1}{2} v g_1
\end{equation}
\addtocounter{equation}{-1}
\renewcommand{\theequation}{\arabic{equation}\mbox{b}}
\begin{equation}
M_{Z} = \frac{1}{2} v \sqrt{g_1^2  + g_2^2}
\end{equation}
and the electric charge $e$ is given in terms of the SU(2) and U(1) gauge couplings, \(g_1\) and \(g_2\) , respectively, by the relation:

\addtocounter{equation}{-1}
\renewcommand{\theequation}{\arabic{equation}\mbox{c}}
\begin{equation}
e = \frac{g_1 g_2}{\sqrt{g_1^2  + g_2^2}} \enskip .
\end{equation}
\\
We perform our calculations in the on-shell renormalization scheme and the physical parameters of the theory are the fermion masses, the Higgs masses, the gauge bosons masses, the angles $\alpha$ and $\beta$, the CKM matrix elements and the electric charge, e. In this scheme, the Weinberg angle is not an independent parameter but just a shorthand notation for the ratio of the W and Z masses, i.e., \(\cos \theta_W = M_W/M_Z\). As was stated and explained by several authors \cite{Hollik} an alternative scheme, which takes advantage of the good precision of the measurements of the Fermi coupling constant, \(G_F\), is obtained replacing \(M_W\) by \(G_F\).

The  introduction of  \(\cal {L_{GF}}\), which essentially removes the contribution of equivalent orbits in the Feynman path integral, induces the existence of ghost fields. After Becchi-Rouet-Stora-Tyutin \cite{BRS74} symmetry was discovered the best way to introduce the ghost contribution is to follow the method advocated by Baulieu \cite{B85}, where this symmetry is promoted to the role of replacing at quantum level the classical gauge symmetry. In this way, one can be sure to obtain all ghost interaction terms and in particular the 4-point interactions${}^{1}$. However, with our choice of gauge fixings, one could also use the better known Faddeev-Popov prescription \cite{FP67}. In any way we obtain:

\footnotetext[1]{see ref. \cite{B85} for a further discussion of this point}
\renewcommand{\theequation}{\arabic{equation}}
\begin{eqnarray}
{\cal L_{FP}} & = & -\overline{C}^+
\left[
\partial^2 + M_W^2
\right] C^- - \overline{C}^-
\left[
\partial^2 + M_W^2
\right] C^+ - \overline{C}^Z
\left[
\partial^2 + M_Z^2
\right] \overline{C}^Z \\
& &  - C^A \enskip \partial^2 \enskip C^A + cubic \quad and \quad quartic \quad terms .\nonumber
\end{eqnarray}
The cubic and quartic terms are similar to the ones in the standard model with the replacement \(H(SM) \rightarrow  H \cos \delta - h \sin \delta \).

\section{The renormalization program}
\subsection{Renormalization of the fields and parameters}

So far, the fields and parameters in the quantum lagrangean are bare. When this lagrangean is used to calculate the Green's functions in pertur\-ba\-tion theory, renormalized fields and couplings have to be introduced. In fact, the calculations of some Feynman diagrams give divergent results. The use of a regularization prescription, in our case dimensional regularization, isolates the divergences in a well prescribed way. Furthermore, the proof of renormalizability, already obtained in 1971 \cite{Ren}, shows that these ultraviolet divergencies can be absorbed by a suitable scaling of the fields and pa\-ram\-eters of the theory. Deciding on a renormalization scheme, in our case the on-shell scheme, fixes the relation between renormalized and unrenormalized Green's functions. This is the general framework for the renormalization of 2HDM that we use. However, even in the simpler standard one-Higgs model, the same on-shell renormalization scheme can be implemented essentially in two ways. In the first one, followed by B\"ohm et al. \cite{BHS87} the gauge boson field renormalization, respects the original gauge symmetry, i.e., the scaling is

\[W^a_{\mu} \rightarrow Z^{1/2}_W W^a_{\mu}\]
\[B_{\mu} \rightarrow Z^{1/2}_B B_{\mu}.\] 
The second alternative followed by Aoki et al. \cite{Aoki} introduces the scaling at the level of the physical fields, W, Z and A. Then, since Z and A have the same quantum numbers they get mixed under renormalization, i.e.,
\renewcommand{\theequation}{\arabic{equation}\mbox{a}}
\begin{equation}
\left[
\begin{array}{c}
Z_{\mu} \\ A_{\mu}
\end{array} \right]_0
= \left[
\begin{array}{cc}
Z_{ZZ}^{1/2} & Z_{ZA}^{1/2} \\
Z_{AZ}^{1/2} & Z_{AA}^{1/2}
\end{array} \right] 
\left[
\begin{array}{c}
Z_{\mu} \\ A_{\mu}
\end{array} \right]
\end{equation}
\addtocounter{equation}{-1}
and
\renewcommand{\theequation}{\arabic{equation}\mbox{b}}
\begin{equation}
W_{\mu 0}^{\pm}=Z_W^{1/2} W_{\mu}^{\pm}
\end{equation}
where the bare fields are denoted by a zero subscript. At first glance it looks as though the first alternative is more economical. However, this is misleading since in this scheme the gauge fixing involves 6 renormalization parameters, whereas in the second, the \(\cal L_{GF}\) is, essentially, unrenormalized. Leaving aside the fermionic sector, the comparison between the renormalization parameters in the two schemes is shown in table I.

In our extension to the 2HDM we found that the second scheme turned out to be the most convenient one. This we will explain in the following paragraph. To close this section let us define some of the entries in table I, in particular the ones that will be used later. The mass counterterms are introduced in the renormalized lagrangean via the scaling

\renewcommand{\theequation}{\arabic{equation}\mbox{a}}
\begin{equation}
M_W^2 \rightarrow M_W^2 + \delta M_W^2
\end{equation}
\addtocounter{equation}{-1}
\renewcommand{\theequation}{\arabic{equation}\mbox{b}}
\begin{equation}
M_Z^2 \rightarrow M_Z^2 + \delta M_Z^2
\end{equation}
\addtocounter{equation}{-1}
\renewcommand{\theequation}{\arabic{equation}\mbox{c}}
\begin{equation}
M_H^2 \rightarrow M_H^2 + \delta M_H^2 \enskip .
\end{equation}
The scaling of the Higgs field and of the would-be Goldstone bosons, i.e.,

\renewcommand{\theequation}{\arabic{equation}\mbox{a}}
\begin{equation}
H \rightarrow Z_H^{1/2} H
\end{equation}
\addtocounter{equation}{-1}
\renewcommand{\theequation}{\arabic{equation}\mbox{b}}
\begin{equation}
G_0 \rightarrow Z_{G_0}^{1/2} G_0
\end{equation}
\addtocounter{equation}{-1}
\renewcommand{\theequation}{\arabic{equation}\mbox{c}}
\begin{equation}
G^+ \rightarrow Z_{G^+}^{1/2} G^+
\end{equation}
introduces the remaining wave function renormalization parameters. The counterterm T, which stands for tadpoles is needed to cancel the one-particle irreducible Green's functions. Later on we will come back to this point. 

\subsection{Renormalization of the gauge fixing}

We start this discussion with the standard one-Higgs model. In the scalar part of the lagrangean, \(\cal L_{S}\), after the symmetry breaking, two-particle mixed terms of the form \(i M_W \partial^{\mu} W_{\mu}^- G^+\) are generated. To define the propagators of the theory those terms have to be eliminated. This is obvious in the unitary gauge where the would-be Goldstone bosons disappear, but it is also true in the \(R_{\xi}\) gauges where the last term in eq. (16) gives a contribution with the opposite sign to the term that we have considered. Clearly, if the gauge fixing is renormalized, the introduction of the same relations between bare and renormalized fields both in  \(\cal L_{S}\) and  \(\cal L_{GF}\) makes this cancellation true to all orders in perturbation theory. Then one is left with no counterterm to renormalize the mixed  \(W_{\mu}^- G^+\) two-particle Green's functions, represented in fig 1.

For illustrative purpose let us write a linear  \(\cal L_{GF}\) in the general form

\renewcommand{\theequation}{\arabic{equation}}
\begin{equation}
{\cal L_{GF}}  = -\frac{1}{\xi}
\left(
\partial^{\mu} W_{\mu}^+ + \xi X^+ G^+
\right)
\left(
\partial^{\mu} W_{\mu}^- - \xi X^- G^-
\right) + \quad ...
\end{equation}
where \(X^+ G^+ \) is defined by the integral

\begin{equation}
X^+ . G^+  = \int d^4 y X ^+ (x-y) G^+ (y)
\end{equation}
and \( X ^+ (x-y) \) is a distribution.

The renormalization implies

\renewcommand{\theequation}{\arabic{equation}\mbox{a}}
\begin{equation}
\xi_W \rightarrow Z_{\xi} \xi_W
\end{equation}
\addtocounter{equation}{-1}
\renewcommand{\theequation}{\arabic{equation}\mbox{b}}
\begin{equation}
W_{\mu} \rightarrow Z_W^{1/2} W_{\mu}
\end{equation}
\addtocounter{equation}{-1}
\renewcommand{\theequation}{\arabic{equation}\mbox{c}}
\begin{equation}
G^+ \rightarrow Z_{G^+}^{1/2} G^+ \enskip .
\end{equation}
Thus, if one renormalizes the function \(X^+\) such that

\addtocounter{equation}{-1}
\renewcommand{\theequation}{\arabic{equation}\mbox{d}}
\begin{equation}
X^+ = Z_{W}^{-1/2} Z_{G^+}^{-1/2} X_R^+
\end{equation}
it is clear that the mixed terms remain unrenormalized. Furthermore, with the condition \(Z_{\xi}=Z_W\), all the terms in the lagrangean given by eq. (23) remain unchanged. However, if one tries to apply the same recipe for the 2HDM we end up with the following counterterms generated by \(\cal L_{GF}\) ,

\renewcommand{\theequation}{\arabic{equation}}
\begin{equation}
{\cal L_{GF}}^{c.t.} = ... + 
\left(
i M_W Z_{G}^{-1/2} Z_{GH}^{1/2} \partial^{\mu} W_{\mu}^+ H^- + h.c. 
\right) \enskip .
\end{equation}                   
Such a counterterm with the opposite sign is generated by the scalar piece of the classical lagrangean, \(\cal L_{S}\), which means that, now, the two-particle WH Green's function is left without counterterm. Fortunately, Baulieu \cite{B85} has proved within the BRST framework that a linear gauge fixing term is not affected by radiative corrections. So, rather than struggling with gauge fixing lagrangeans with extra \(\xi\) parameters, we will follow Ross and Taylor \cite{RT73} in their celebrated paper and do not renormalize \(\cal L_{GF}\) given by eq (16). In other words, the fields and parameters in this eq. are already assumed to be the renormalized ones. Furthermore, in the calculation we choose \(\xi_A = \xi_Z = \xi_W = 1\), which corresponds to the usual Feynman-t'Hooft gauge.

\subsection{One-particle irreducible Green's functions}

After the discovery of the BRST symmetry, the renormalization of gauge theories is proved using BRST Ward identities. In the one-doublet standard model, these identities are independent of the sign of the $ \mu^2$ term in the Higgs potential. Then, the proof of the renormalizability of the spontaneously broken standard theory, follows immediately.

Recently \cite{SN94}, Schilling and van Nieuwenhuizen have explicitly proved the multiplicative renormalization of an SU(2) gauge model. In this case, both the vacuum expectation value, $v$, and the scalar field are multiplicatively renormalized by a different Z factor. Hence, it is clear that, in this case, the tree level condition $-\mu^2 + \lambda v^2 = 0$ is not mantained in higher orders. In the potential, $-\mu^2 + \lambda v^2$ is the coefficient of the term linear in the Higgs field. So, in this multiplicative renormalization scheme there will be renormalized linear terms in H.

An alternative is to introduce an additive renormalization scheme for the scalar fields. In other words, we shift the fields by an additive constant such that their vacuum expectation value vanish order by order. This is the scheme that we follow here.

In fig.2 we show these so-called tadpole diagrams together with their counterterms chosen in such a way that the renormalized Green's functions vanish.   
These conditions, namely

\renewcommand{\theequation}{\arabic{equation}\mbox{a}}
\begin{equation}
\Sigma_H + T_H = 0
\end{equation}
\addtocounter{equation}{-1}
\renewcommand{\theequation}{\arabic{equation}\mbox{b}}
\begin{equation}
\Sigma_h + T_h = 0
\end{equation}
fix, order by order, the values of \(T_{H, h}\). Notice that, because of CP con\-ser\-vation, there is no tadpole diagram for the pseudoscalar field.

Naively one could assume that this corresponds simply to forget about the tadpole diagrams. Indeed, this is the case, for any diagram that differs from a lower order one by a simple addition of a tadpole subgraph. However, we still have to evaluate the counterterms given by eqs. (27,a,b) because those counterterms are going to influence the results for two-point renormalized Green's functions. This is already seen in eq. (10) and it will be shown in the next paragraph.

\subsection{Two-particle irreducible Green's functions}

In this section we discuss the renormalization of the two-point Green's func\-tions. The only differences from the standard model are in the scalar sector and in the mixing between the scalar and gauge boson sectors. Hence, we only discuss those cases and refer to Aoki \cite{Aoki} for the remaining two-point functions.

Let us start by showing that the bilinear scalar terms in the tree level La\-grangean have exactly the same form that in the one-Higgs SM Lagrangean. Using again only the charged sector as an example, the bilinear terms of the kinetic Lagrangean can be written as

\renewcommand{\theequation}{\arabic{equation}}
\begin{eqnarray}
\sum_{i}
\left(
D_{\mu}\phi_{i}
\right)^\dagger
\left(
D^\mu \phi_{i}
\right) & = & 
\sum_{i}
\left(
\partial_{\mu}a_{i}^+
\right)
\left(
\partial^\mu a_{i}^-
\right) + M_W^2 x_i^2 W_{\mu}^+ W^{\mu -} \\ & & \nonumber \\
& & + \enskip i M_W 
\left(
W_{\mu}^+ (\partial^{\mu} a_{i}^-) - W^{\mu -} (\partial_{\mu} a_{i}^+) 
\right) + ... \nonumber
\end{eqnarray}
where the two fields \(a_i\) are eigenstates of SU(2), \(x_1 = v_1/v = \cos \beta\),  \(x_2 =\) \(v_2/v = \sin \beta\) and  \(v^2 = v_1^2 + v_2^2\).
The scalars kinetic term is

\begin{equation}
\left(
\partial_{\mu}a_{1}^+
\right)
\left(
\partial^\mu a_{1}^-
\right) +
\left(
\partial_{\mu}a_{2}^+
\right)
\left(
\partial^\mu a_{2}^-
\right)
\end{equation}
and the W boson mass term is

\begin{equation}
M_W^2 W_{\mu}^+ W^{\mu -}
\left(
x_1^2 + x_2^2
\right)
\end{equation}
and finally the mixing term is

\begin{equation}
i M_W \partial^{\mu} W_{\mu}^+
\left(
x_1 a_{1}^- + x_2 a_{2}^- 
\right) + h.c. \enskip .
\end{equation}
As we have seen, the relations between the SU(2) eigenstates ($a_1$ and $a_2$) and the mass eigenstates (H and G) is:

\begin{equation}
\left\{
\begin{array}{l}
a_{1}^-  = - x_2 H^- + x_1 G^- \\
a_{2}^-  =  x_1 H^- + x_2 G^-
\end{array}
\right. \enskip .
\end{equation}
So we readily see that only the Goldstone boson, $x_1 a_1^- + x_2 a_2^- = G^-$, appears in eq. (31) which means that there are no extra terms in the mixing. On the other hand, the terms (29) and (30) can be written in the following form

\begin{equation}
\left(
x_1^2 + x_2^2
\right)
\left[
(\partial^{\mu} H^+)(\partial_{\mu} H^-) + (\partial^{\mu} G^+)(\partial_{\mu} G^-) + M_W^2 W_{\mu}^+ W^{\mu - } 
\right]
\end{equation}
Now, if we renormalize the angle with the condition \(\beta_0 = \beta + \delta \beta\), we get

\begin{equation}
\left(
x_1^2 + x_2^2
\right) = \cos^2 (\beta + \delta \beta) + \sin^2 (\beta + \delta \beta) = 1
\end{equation}
and, of course, this relation holds to any order of perturbation theory.

We can now start the renormalization program from the tree-level Lagrangean. The renormalized fields and masses are defined by the relations

\begin{eqnarray}
\left[
\begin{array}{l}
H^{\pm} \\ G^{\pm}
\end{array} \right]_0
& = & \left[
\begin{array}{cc}
Z_{H^+ H^+}^{1/2} & Z_{H^+ G^+}^{1/2} \\
Z_{G^+ H^+}^{1/2} & Z_{G^+ G^+}^{1/2}
\end{array} \right] 
\left[
\begin{array}{l}
H^{\pm} \\ G^{\pm}
\end{array} \right]  \\ \nonumber \\ \nonumber \\
& & M_{H^+ 0}^2 = M_{H^+}^2 + \delta M_{H^+}^2 \nonumber \enskip .
\end{eqnarray}
Now we have to find the counterterms for the two point functions. The bilinear terms in the Lagrangean for the charged Higgs sector are
      
\begin{eqnarray}
{\cal L} & = & - H^+
\left[
Z_{H^+ H^+} (\partial^2 + M_{H^+}^2 + \delta M_{H^+}^2) \right. \nonumber \\
& & \left. + Z_{G^+ H^+} \partial^2 + 2 \frac{T_{\alpha \beta}+T_{\delta} \cos^2 \beta}{v \sin 2 \beta}
\right] H^-  \nonumber \\
& & - G^+
\left[
Z_{H^+ G^+} (\partial^2 + M_{H^+}^2 + \delta M_{H^+}^2)  \right. \\
& & \left. + Z_{G^+ G^+} \partial^2 + 2 \frac{T_{\alpha \beta}+ T_{\delta}\sin^2 \beta}{v \sin 2 
\beta}
\right] G^-  \nonumber \\
& & - H^+
\left[
Z_{H^+ H^+}^{1/2} Z_{H^+ G^+}^{1/2} (\partial^2 + M_{H^+}^2 + \delta M_{H^+}^2) \right. \nonumber \\
& & + \left. Z_{G^+ G^+}^{1/2} Z_{G^+ H^+}^{1/2} \partial^2 + \frac{T_{\delta}}{v}
\right] G^- + h.c. \enskip . \nonumber
\end{eqnarray}
Using the usual recipe for on-shell renormalization, that is, demanding that the pole stays at the physical mass and that the residue is one, we arrive at the following set of renormalization conditions

\renewcommand{\theequation}{\arabic{equation}\mbox{a}}
\begin{equation}
\Sigma_{H^+ H^+} ( M_{H^+}^2) - Z_{H^+ H^+} \delta M_{H^+}^2+ Z_{G^+ H^+} M_{H^+}^2 - 2 \frac{T_{\alpha \beta}+ T_{\delta} \cos^2 \beta}{v \sin 2 \beta} = 0
\end{equation}
\addtocounter{equation}{-1}
\renewcommand{\theequation}{\arabic{equation}\mbox{b}}
\begin{equation}
\frac{d}{dq^2} \Sigma_{H^+ H^+} ( M_{H^+}^2) + Z_{H^+ H^+} + Z_{G^+ H^+} = 0
\end{equation}
\\

\renewcommand{\theequation}{\arabic{equation}\mbox{a}}
\begin{equation}
\Sigma_{G^+ G^+} ( 0) - Z_{H^+ G^+} (M_{H^+}^2 + \delta M_{H^+}^2) - 2 \frac{T_{\alpha \beta}+ T_{\delta}\sin^2 \beta}{v \sin 2 \beta} = 0
\end{equation}
\addtocounter{equation}{-1}
\renewcommand{\theequation}{\arabic{equation}\mbox{b}}
\begin{equation}
\frac{d}{dq^2} \Sigma_{G^+ G^+} (0) + Z_{G^+ G^+} + Z_{H^+ G^+} = 0
\end{equation}
\\
\renewcommand{\theequation}{\arabic{equation}\mbox{a}}
\begin{equation}
\Sigma_{H^+ G^+} (0) - Z_{H^+ H^+}^{1/2} Z_{H^+ G^+}^{1/2} (M_{H^+}^2 + \delta M_{H^+}^2) - \frac{T_{\delta}}{v} = 0
\end{equation}
\addtocounter{equation}{-1}
\renewcommand{\theequation}{\arabic{equation}\mbox{b}}
\begin{equation}
\Sigma_{G^+ H^+} (M_{H^+}^2) - Z_{H^+ H^+}^{1/2} Z_{H^+ G^+}^{1/2} \delta M_{H^+}^2+Z_{G^+ G^+}^{1/2} Z_{G^+ H^+}^{1/2}M_{H^+}^2 - \frac{T_{\delta}}{v} = 0.
\end{equation}
\\
With these six equations we can determine the five renormalization con\-stants. Notice the explicit appearance of the tadpole counterterms. There is one dependent equation due to a Ward identity in the charged sector, which is:

\renewcommand{\theequation}{\arabic{equation}}
\begin{equation}
\begin{array}{l}
<0|T \partial^{\mu} W_{\mu}^+  \partial^{\nu} W_{\nu}^-|0> - i M_W <0|T G^+ \partial^{\nu} W_{\nu}^-|0> \\ \\ 
\enskip + i M_W <0|T \partial^{\mu} W_{\mu}^+ G^-|0> + M_W^2 <0|T G^+ G^-|0> = 0 \enskip .
\end{array}
\end{equation}
Finally let us discuss the mixed terms in the charged sector. Bearing in mind the discussion about the gauge fixing Lagrangean in the previous section, the counterterms can be taken from
	      
\begin{equation}
\begin{array}{l}
{\cal L}  = i (M_{W}^2 + \delta M_{W}^2)^{1/2} Z_W^{1/2}Z_{G^+ G^+}^{1/2}
W_{\mu}^- \partial^{\mu} G^+ + h.c.
\\ \\ \quad
+ i (M_{W}^2 + \delta M_{W}^2)^{1/2} Z_W^{1/2}Z_{G^+ H^+}^{1/2}
W_{\mu}^- \partial^{\mu} H^+ + h.c. \enskip .
\end{array}
\end{equation}
The gauge fixing Lagrangean (23) will cancel the tree level terms in (41) and so, the final mixed Lagrangean is, in fact, a counterterm Lagrangean for the self-energies WG and WH. Notice that we did not explicitly introduce any counterterms for the Green's functions WG and WH. So, we end up this section by writing simbolically $Z_{W^+ G^+}$ and $Z_{W^+ H^+}$ as

\begin{equation}
Z_{W^+ G^+}^{1/2}  = i k_{\mu}(M_{W}^2 + \delta M_{W}^2)^{1/2} Z_W^{1/2}Z_{G^+ G^+}^{1/2}
\end{equation}
\begin{equation}
Z_{W^+ H^+}^{1/2}  = i k_{\mu}(M_{W}^2 + \delta M_{W}^2)^{1/2} Z_W^{1/2}Z_{G^+ H^+}^{1/2}
\end{equation}
The complete set of counterterms for the scalar and mixed sectors can be found in Appendix B.

\subsection{Three-particle irreducible Green's functions}

In the on-shell renormalization scheme that we have adopted, the gauge couplings $g_1$ and $g_2$ are not independent parameters. In fact, they are both related to the gauge boson masses and to the electric charge, e, i.e.,

\begin{eqnarray}
& & g_1  = e \frac{M_Z}{M_W} \nonumber \\
& & g_2  = e \frac{M_Z}{(M_W^2-M_Z^2)^{1/2}} \enskip . \nonumber
\end{eqnarray}
Then, in the one-Higgs model, only one further renormalization constant $Y=\delta e/e$ remains to be fixed. This is simply done by imposing the condition

\begin{equation}
\overline{u}(m_f) \Gamma_R^{\mu}u(m_f) |_{k^{\mu} \rightarrow 0} =  \overline{u} \gamma^{\mu}u
\end{equation}
for any charged fermion, where $ \Gamma_R^{\mu}$ is the renormalized three-point photon fermion vertex. Usually, following the traditional QED prescription, where the Thompson limit was introduced to define $\alpha=e^2/(4 \pi)$, one uses the electron as the charged fermion. However, the universality of the on-shell charge, guarantees that one can use any charged fermion. Since the theory is by itself well defined, one could alternatively fix Y by using the renormalized $W^+ W^- \gamma$ \enskip three-point function, namely

\begin{eqnarray}
& & \left[ \epsilon_{\beta}(p) \epsilon_{\gamma}(q) \Gamma_R^{\beta \gamma \mu} \enskip  \right] _{k^{\mu} \rightarrow 0 \enskip p^2=q^2=M_W^2} =  \lim_{k^{\mu} \rightarrow 0}  \left[\epsilon(q).(k-p) \epsilon^{\mu}(p) \right. \nonumber \\  \\
& & \left. + (p-q)^{\mu}\epsilon(p). \epsilon(q) + (q-k).p \epsilon^{\mu}(q)
\right]=0 \enskip . \nonumber
\end{eqnarray}

Besides the gauge coupling renormalization, fixed by the photon coup\-ling, the W quark-quark vertex requires the additional renormalization of the Cabibbo-Kobayashi-Maskawa (CKM) matrix. For the standard one-Higgs model this renormalization of the CKM matrix was evaluated by Den\-ner and Sack \cite{DS90}. In this article we extend this analysis to the 2HDM.

Let us consider the decay $W^+ \rightarrow u_I \overline{d}_j$, where I, j= 1, 2, 3 are the gen\-er\-a\-tion indices (upper case for up quarks). At tree-level the decay amplitude is

\begin{equation}
T=V_{Ij}T_o
\end{equation}
with
\begin{equation}
T_o = -\frac{g}{\sqrt{2}} \overline{u}_I (m_I) \not\epsilon \gamma_L v (m_j).
\end{equation}
At one-loop, in the on-shell renormalization scheme the self-energy cor\-rec\-tions to the external legs vanish and the proper vertex diagrams give an amplitude $T_1^v$ that can be written in the form

\begin{equation}
T_1^v = V_{Ij} T_o \Delta ,
\end{equation}
where $\Delta$ stands for the result of the loop calculation. To obtain the full one-loop amplitude one has to add the counterterms, i.e.,

\begin{equation}
T_1=T_1^v + T_1^c
\end{equation}
with
\begin{eqnarray}
T_1^c & = & V_{Ij} T_o \left[ \frac{\delta g}{g} - \frac{1}{2} \delta Z_W \right] + \frac{1}{2} T_o \left[ \sum_{J} \delta Z_{JI}^{*L} V_{Jj} +  \sum_{i} V_{Ii} \delta Z_{ij}^{L} \right] \\
& & + T_o \delta V_{Ij} \nonumber
\end{eqnarray}
Now we have to face the problem of imposing some conditions to fix the CKM counterterms $\delta V_{Ij}$. Denner and Sack \cite{DS90} have split the quark wave-function renormalization parameters, $\delta Z^L$ into its hermitian and antihermitian con\-tri\-butions, namely,

\begin{equation}
\delta Z^L = \frac{1}{2}(\delta Z^L + \delta Z^{*L}) +  \frac{1}{2}(\delta Z^L - \delta Z^{*L}) 
\end{equation}
and then they have fixed $\delta V_{Ij}$ by the condition

\begin{equation}
\delta V_{Ij} = - \frac{1}{4} \left[ \sum_{J}( \delta Z^{*L} - \delta Z^L)_{JI}V_{Jj} +  \sum_{i} V_{Ii} ( \delta Z^{L} - \delta Z^{*L})_{ij} \right] .
\end{equation}
It is possible to prove \cite{DS90} that $\delta V_{Ij}$ is needed precisely to cancel the divergent contribution to the righthand side of eq. (52). Hence, the use of eq. (52) to fix also the finite piece of $\delta V_{Ij}$ is a possible choice. Alternatively, one could select four physical $Wq \overline{q}$ decay processes and impose the vanishing of $T_1$ for these decays. In this case, the transitions $W^+ \rightarrow u \overline{d}$; $W^+ \rightarrow u \overline{s}$; $W^+ \rightarrow u \overline{b}$ and $t \rightarrow b W^+$ could form an interesting set. However, this process has the clear disavantage of shifting all one-loop correction to some amplitudes.

Since the renormalization of the CKM matrix vanish in the limit of degenerate down quark masses, most loop corrections to the W decay process are done in this approximation. This is equivalent to drop the last term in eq. (50) and, in the same term, to replace the sum over J and i simply by the J=I and j=i contributions. Hence, in this approximation $T_1$ is directly proportional to a single CKM element, $V_{Ij}$. As far as we know all standard model analysis of the values of the CKM matrix elements are done in this approximation. In fact, the work of Denner and Sack has shown that the error of this approximation is of the order $10^{-6}$, far smaller than any other theoretical and experimental uncertainties.

In the 2HDM, one can do a similar analysis with the difference that there are further contributions to the irreducible vertex and to $\delta Z_L$ coming from diagrams with neutral and charged Higgs. Because some of these vertices could be enhanced by the factor $\tan \beta$($\cot \beta$), one could expect to see such enhancement in the result.

In the 2HDM there are two further couplings, $\alpha$ and $\beta$, that need to be renormalized. This can be done imposing some physical conditions on the renormalized three-point or four-point scalar vertex functions. There are in this model 8 cubic and 14 quartic vertices among the neutral and charged Higgs and any two of those can be selected. However, most of these vertices have a complicated dependence on the angles and, furthermore, without knowing the Higgs masses it is difficult to select a physical process like for instance $H \rightarrow hh$. Luckily, the vertices $\overline{e}eh$ and $H^{\pm} e \nu$ which induce the tree-level decays $h \rightarrow e^+ e^-$ and $H^- \rightarrow e^- \overline{\nu}$, have a simple dependence on the angles (see table II) and, at the same time, we already know that the present bounds on the Higgs masses allow these decays to occur.

In a recent calculation \cite{rui2} of the top-loop contribution to the decay $H^+ \rightarrow h W^+$, where the vertex depends only on the combination $\beta - \alpha$, we renormalize  $(\beta - \alpha)$ using the corresponding process $H^+ \rightarrow H W^+$. In the absence of any information on Higgs scattering and Higgs leptonic decays this is perhaps the only consistent way to proceed.
\bigskip
\bigskip
\bigskip

\textbf{Acknowledgements}

We thank J.C. Rom\~ao and J.P. Silva for a critical reading of the manuscript.

\newpage
\appendix
\section{Feynman rules}
In this appendix we present the Feynman rules for the interactions involving scalar fields. All other interactions are standard and can be found in \cite{Aoki}. We have chosen the Feynman-t'Hooft gauge and followed the convention that all the momenta in the vertices are incoming.

We start by defining the following quantities:
\begin{eqnarray}
A_{\alpha \beta} \equiv \cos^3 \beta \sin \alpha + \sin^3 \beta \cos \alpha \nonumber \\
B_{\alpha \beta} \equiv \cos^3 \beta \cos \alpha + \sin^3 \beta \sin \alpha \nonumber \\
C_{\alpha \beta} \equiv \sin^3 \alpha \cos \beta + \cos^3 \alpha \sin \beta \nonumber \\
D_{\alpha \beta} \equiv \cos^3 \alpha \cos \beta - \sin^3 \alpha \sin \beta \enskip . \nonumber \\ \nonumber
\end{eqnarray}
In the Yukawa lagrangean, the fermions can couple with the scalars in four different and independent ways, with no flavour changing. The couplings for those models are shown in table II. In Model I only $\phi_2$ couples to all fermions; in Model II $\phi_2$ couples to the quarks and $\phi_1$ coulpes to the leptons; in Model III $\phi_2$ couples to the up quarks and to the leptons and $\phi_1$ couples to the down quarks; finally in Model IV $\phi_2$ couples to the up quarks and $\phi_1$ couples to the down quarks and the leptons.
The two Higgs doublets $\phi_1$ and $\phi_2$ are defined in the expression (4). These couplings will be used in section A.3.

\newpage
\subsection{Triple scalar vertices}

\begin{tabular}{lcl}
$H^+ H^- h$ & & $-\frac{ig}{M_W} \left( \frac{M_h^2}{\sin 2 \beta} B_{\alpha \beta}-M_{H^+}^2 \sin \delta \right)$ \\ \\
$H^+ H^- H$ & & $-\frac{ig}{M_W} \left( \frac{M_H^2}{\sin 2 \beta} A_{\alpha \beta}+M_{H^+}^2 \cos \delta \right)$ \\ \\
$A A h$ & & $-\frac{ig}{M_W} \left( \frac{M_h^2}{\sin 2 \beta} B_{\alpha \beta}-M_{A}^2 \sin \delta \right)$ \\ \\
$A A H$ & & $-\frac{ig}{M_W} \left( \frac{M_H^2}{\sin 2 \beta} A_{\alpha \beta}+M_{A}^2 \cos \delta \right)$ \\ \\
$h h h$ & & $\frac{3ig}{M_W} \frac{M_h^2}{\sin 2 \beta} D_{\alpha \beta}$ \\ \\
$H H H$ & & $-\frac{3ig}{M_W} \frac{M_H^2}{\sin 2 \beta} C_{\alpha \beta}$ \\ \\
$h H H$ & & $\frac{ig}{2M_W} \frac{\sin 2 \alpha \sin \delta (2 M_H^2 + M_h^2)}{\sin 2 \beta}$ \\ \\
$h h H$ & & $-\frac{ig}{2M_W} \frac{\sin 2 \alpha \cos \delta (M_H^2 + 2 M_h^2)}{\sin 2 \beta}$ \\ \\
$h H^{\mp}G^{\pm}$ & & $-\frac{ig}{2M_W} \cos \delta (M_h^2 - M_{H^+}^2)$ \\ \\
$H H^{\mp}G^{\pm}$ & & $\frac{ig}{2M_W} \sin \delta (M_H^2 - M_{H^+}^2)$ \\ \\
$h A G_0$ & & $-\frac{ig}{2M_W} \cos \delta (M_h^2 - M_{A}^2)$ \\ \\
\end{tabular}

\newpage
\begin{tabular}{lcl}
$H A G_0$ & & $\frac{ig}{2M_W} \sin \delta (M_H^2 - M_{A}^2)$ \\ \\
$h G_0 G_0$ & & $\frac{ig}{2M_W} \sin \delta M_h^2$ \\ \\
$h G^+ G^+$ & & $\frac{ig}{2M_W} \sin \delta M_h^2$ \\ \\
$H G_0 G_0$ & & $-\frac{ig}{2M_W} \cos \delta M_H^2$ \\ \\
$H G^+ G^+$ & & $-\frac{ig}{2M_W} \cos \delta M_H^2$ \\ \\
$A H^{\mp}G^{\pm}$ & & $\pm \frac{g}{2M_W} (M_A^2 - M_{H^+}^2)$ \\ \\
\end{tabular}

\subsection{Quartic scalar vertices}

\begin{tabular}{lcl}
$H^+ H^- H^+ H^-$ & & $-\frac{ig^2}{\sin^2 2 \beta M_W^2}(M_H^2 A_{\alpha \beta}^2 + M_h^2 B_{\alpha \beta}^2)$ \\ \\
$A A A A$ & & $-\frac{3ig^2}{\sin^2 2 \beta M_W^2}(M_H^2 A_{\alpha \beta}^2 + M_h^2 B_{\alpha \beta}^2)$ \\ \\
$A A H^+ H^-$ & & $-\frac{ig}{\sin^2 2 \beta M_W^2}(M_H^2 A_{\alpha \beta}^2 + M_h^2 B_{\alpha \beta}^2)$ \\ \\
$H^+ H^- h h$ & & $-\frac{ig^2}{2 M_W^2}\left[ \frac{1}{\sin^2 2 \beta}(M_H^2 A_{\alpha \beta} \sin 2 \alpha \cos \delta - 2 M_h^2 B_{\alpha \beta}D_{\alpha \beta}) + 2 M_{H^+}^2 \sin^2 \delta \right]$ \\ \\
\end{tabular}

\newpage
\begin{tabular}{lcl}
$H^+ H^- H H$ & & $-\frac{ig^2}{2 M_W^2}\left[ \frac{1}{\sin^2 2 \beta}(2 M_H^2 A_{\alpha \beta} C_{\alpha \beta} + M_h^2 B_{\alpha \beta}\sin 2 \alpha \sin \delta) + 2 M_{H^+}^2 \cos^2 \delta \right]$ \\ \\
$A A h h$ & & $-\frac{ig^2}{2 M_W^2}\left[ \frac{1}{\sin^2 2 \beta}(M_H^2 A_{\alpha \beta} \sin 2 \alpha \cos \delta - 2 M_h^2 B_{\alpha \beta}D_{\alpha \beta}) + 2 M_{A}^2 \sin^2 \delta \right]$ \\ \\
$A A H H$ & & $-\frac{ig^2}{2 M_W^2}\left[ \frac{1}{\sin^2 2 \beta}(2 M_H^2 A_{\alpha \beta} C_{\alpha \beta} + M_h^2 B_{\alpha \beta}\sin 2 \alpha \sin \delta) + 2 M_{A}^2 \cos^2 \delta \right]$ \\ \\
$H^+ H^- H h$ & & $-\frac{ig^2}{2 M_W^2}\left[ \frac{1}{\sin^2 2 \beta}(M_H^2 A_{\alpha \beta} \sin 2 \alpha \sin \delta + M_h^2 B_{\alpha \beta}\sin 2 \alpha \cos \delta) - M_{H^+}^2 \sin 2 \delta \right]$ \\ \\
$A A H h$ & & $-\frac{ig^2}{2 M_W^2}\left[ \frac{1}{\sin^2 2 \beta}(M_H^2 A_{\alpha \beta} \sin 2 \alpha \sin \delta + M_h^2 B_{\alpha \beta}\sin 2 \alpha \cos \delta) - M_{A}^2 \sin 2 \delta \right]$ \\ \\
$h h h h$ & & $-\frac{3ig^2}{4 \sin^2 2 \beta M_w^2} (4 M_h^2 D_{\alpha \beta}^2 + M_H^2 \sin^2 2 \alpha \cos^2 \delta)$ \\ \\
$H H H H$ & & $-\frac{3ig^2}{4 \sin^2 2 \beta M_W^2} (M_h^2 \sin^2 2 \alpha \sin^2 \delta + 4 M_H^2 C_{\alpha \beta}^2)$ \\ \\
$h h h H$ & & $-\frac{3ig^2}{8 \sin^2 2 \beta M_W^2} (4 M_h^2 D_{\alpha \beta} \sin 2 \alpha \cos \delta + M_H^2 \sin^2 2 \alpha \sin 2 \delta)$ \\ \\
$H H H h$ & & $-\frac{3ig^2}{8 \sin^2 2 \beta M_W^2} (M_h^2 \sin^2 2 \alpha \sin 2 \delta + 4 M_H^2 C_{\alpha \beta} \sin 2 \alpha \sin \delta)$ \\ \\
$h h H H$ & & $- \frac{i g^2 \sin 2 \alpha}{4 \sin 2 \beta M_W^2} \left[ M_H^2 - M_h^2 + \frac{3 \sin 2 \alpha}{sin 2 \beta} (\sin^2 \delta M_H^2 + \cos^2 \delta M_h^2) \right]$ \\ \\
$A A G_0 G_0$ & & $- \frac{ig^2}{4 M_W^2} \left[ \frac{\sin 2 \alpha}{sin 2 \beta}( M_H^2 - M_h^2) + 3 (\sin^2 \delta M_H^2 + \cos^2 \delta M_h^2) \right]$ \\ \\
\end{tabular}

\newpage
\begin{tabular}{lcl}
$H^+ H^- G^+ G^-$ & & $- \frac{ig^2}{4 M_W^2} \left[ M_A^2 + \frac{\sin 2 \alpha}{sin 2 \beta}( M_H^2 - M_h^2) + 3 (\sin^2 \delta M_H^2 + \cos^2 \delta M_h^2) \right]$ \\ \\
$H^{\pm} H^{\pm} G^{\mp} G^{\mp}$ & & $- \frac{ig^2}{2 M_W^2}(- M_A^2 + \sin^2 \delta M_H^2 + \cos^2 \delta M_h^2)$ \\ \\
$H^{\mp} H^{\pm} A G_0$ & & $- \frac{ig^2}{4 M_W^2}(- M_{H^+}^2 + \sin^2 \delta M_H^2 + \cos^2 \delta M_h^2)$ \\ \\
$G^+ G^- A A$ & & $- \frac{ig^2}{2 M_W^2} \left[ M_{H^+}^2 + \frac{1}{\sin 2 \beta}(\cos \delta A_{\alpha \beta}M_H^2 - \sin \delta  B_{\alpha \beta} M_h^2) \right]$ \\ \\
$H^+ H^- G_0 G_0$ & & $- \frac{ig^2}{2 M_W^2} \left[ M_{H^+}^2 + \frac{1}{\sin 2 \beta}(\cos \delta A_{\alpha \beta}M_H^2 - \sin \delta  B_{\alpha \beta} M_h^2) \right]$ \\ \\
$H^+ H^- H^{\mp}G^{\pm}$ & & $- \frac{ig^2}{M_W^2} \left[ \frac{\sin \delta}{\sin 2 \beta}(A_{\alpha \beta}M_H^2 + B_{\alpha \beta} M_h^2) \right]$ \\ \\
$H^+ H^- G_0 A$ & & $- \frac{ig^2}{2 M_W^2} \left[ \frac{\sin \delta}{\sin 2 \beta}(A_{\alpha \beta}M_H^2 + B_{\alpha \beta} M_h^2) \right]$ \\ \\
$A A A G_0$ & & $- \frac{3ig^2}{2 M_W^2} \left[ \frac{\sin \delta}{\sin 2 \beta}(A_{\alpha \beta}M_H^2 + B_{\alpha \beta} M_h^2) \right]$ \\ \\
$A A H^{\mp}G^{\pm}$ & & $- \frac{ig^2}{2M_W^2} \left[ \frac{\sin \delta}{\sin 2 \beta}(A_{\alpha \beta}M_H^2 + B_{\alpha \beta} M_h^2) \right]$ \\ \\
$G^+ G^- G_0 A$ & & $- \frac{ig^2}{8 M_W^2} \sin 2 \delta (M_H^2 - M_h^2)$ \\ \\
$G^+ G^-  H^{\mp}G^{\pm}$ & & $- \frac{ig^2}{4 M_W^2} \sin 2 \delta (M_H^2 - M_h^2)$ \\ \\
\end{tabular}

\newpage
\begin{tabular}{lcl}
$G_0 G_0  G_0 A$ & & $- \frac{3ig^2}{8 M_W^2} \sin 2 \delta (M_H^2 - M_h^2)$ \\ \\
$G_0 G_0  H^{\mp}G^{\pm}$ & & $- \frac{ig^2}{8 M_W^2} \sin 2 \delta (M_H^2 - M_h^2)$ \\ \\
$G^+ G^- h h$ & & $- \frac{ig^2}{4 M_W^2} \left[ \frac{1}{\sin 2 \beta}(\sin 2 \alpha \cos^2 \delta M_H^2 - 2 \sin \delta D_{\alpha \beta} M_h^2) + 2 \cos^2 \delta M_{H^+}^2 \right]$ \\ \\
$G_0 G_0 h h$ & & $- \frac{ig^2}{4 M_W^2} \left[ \frac{1}{\sin 2 \beta}(\sin 2 \alpha \cos^2 \delta M_H^2 - 2 \sin \delta D_{\alpha \beta} M_h^2) + 2 \cos^2 \delta M_{A}^2 \right]$ \\ \\
$G^+ G^- H H$ & & $- \frac{ig^2}{4 M_W^2} \left[ \frac{1}{\sin 2 \beta}(2 \cos \delta C_{\alpha \beta} M_H^2 - \sin^2 \delta \sin 2 \alpha M_h^2) + 2 \sin^2 \delta M_{H^+}^2 \right]$ \\ \\
$G_0 G_0 H H$ & & $- \frac{ig^2}{4 M_W^2} \left[ \frac{1}{\sin 2 \beta}(2 \cos \delta C_{\alpha \beta} M_H^2 - \sin^2 \delta \sin 2 \alpha M_h^2) + 2 \sin^2 \delta M_{A}^2 \right]$ \\ \\
$H^{\mp}G^{\pm}H H$ & & $- \frac{ig^2}{8 M_W^2} \left[ \frac{1}{\sin 2 \beta}(4 \cos \delta C_{\alpha \beta} M_H^2 + \sin 2 \delta \sin 2 \alpha M_h^2) - 2 \sin 2 \delta M_{H^+}^2 \right]$ \\ \\
$H^{\mp}G^{\pm}h h$ & & $- \frac{ig^2}{8 M_W^2} \left[ \frac{1}{\sin 2 \beta}( \sin 2 \delta \sin 2 \alpha M_H^2 + 4 \cos \delta D_{\alpha \beta} M_h^2) + 2 \sin 2 \delta M_{H^+}^2 \right]$ \\ \\
$A G_0 H H$ & & $- \frac{ig^2}{8 M_W^2} \left[ \frac{1}{\sin 2 \beta}(4 \cos \delta C_{\alpha \beta} M_H^2 + \sin 2 \delta \sin 2 \alpha M_h^2) - 2 \sin 2 \delta M_{A}^2 \right]$ \\ \\
$A G_0 h h$ & & $- \frac{ig^2}{8 M_W^2} \left[ \frac{1}{\sin 2 \beta}(\sin 2 \alpha \sin 2 \delta M_H^2 + 4 \cos \delta D_{\alpha \beta} M_h^2) + 2 \sin 2 \delta M_{A}^2 \right]$ \\ \\
$G^{\mp} H^{\pm} h A$ & & $\pm \frac{g^2}{4 M_W^2} \sin \delta (M_A^2 - M_{H^+}^2)$ \\ \\
\end{tabular}

\newpage
\begin{tabular}{lcl}
$G^{\mp} H^{\pm} H G_0$ & & $\pm \frac{g^2}{4 M_W^2} \sin \delta (M_A^2 - M_{H^+}^2)$ \\ \\
$G^{\mp} H^{\pm} h G_0$ & & $\pm \frac{g^2}{4 M_W^2} \cos \delta (M_A^2 - M_{H^+}^2)$ \\ \\
$G^{\mp} H^{\pm} H A$ & & $\mp \frac{g^2}{4 M_W^2} \cos \delta (M_A^2 - M_{H^+}^2)$ \\ \\
$G^+ G^- h H$ & & $- \frac{ig^2}{8 M_W^2} \left[ \frac{\sin 2 \alpha}{\sin 2 \beta}(M_H^2 - M_h^2) + 2 \sin 2 \delta M_{H^+}^2 \right]$ \\ \\
$G_0 G_0 h H$ & & $- \frac{ig^2}{8 M_W^2} \left[ \frac{\sin 2 \alpha}{\sin 2 \beta}(M_H^2 - M_h^2) + 2 \sin 2 \delta M_{A}^2 \right]$ \\ \\
$G^{\mp} H^{\pm} h H$ & & $- \frac{ig^2}{4 M_W^2} \left[ \frac{\sin 2 \alpha}{\sin 2 \beta}(\sin^2 \delta M_H^2 + \cos^2 \delta M_h^2) - \cos 2 \delta M_{H^+}^2 \right]$ \\ \\
$A G_0 h H$ & & $- \frac{ig^2}{4 M_W^2} \left[ \frac{\sin 2 \alpha}{\sin 2 \beta}(\sin^2 \delta M_H^2 + \cos^2 \delta M_h^2) - \cos 2 \delta M_{A}^2 \right]$ \\ \\
$G^+ G^- G^+ G^-$ & & $- \frac{ig^2}{4 M_W^2} (\sin^2 \delta M_h^2 + \cos^2 \delta M_H^2)$ \\ \\
$G_0 G_0 G_0 G_0$ & & $- \frac{3ig^2}{4 M_W^2} (\sin^2 \delta M_h^2 + \cos^2 \delta M_H^2)$ \\ \\
$G^+ G^- G_0 G_0$ & & $- \frac{ig^2}{4 M_W^2} (\sin^2 \delta M_h^2 + \cos^2 \delta M_H^2)$ \\ \\
\end{tabular}

\newpage
\subsection{Fermion-scalar vertices}

\begin{tabular}{lcl}
$\overline{e}_i e_i h$ & & $\frac{ig}{2 M_W}\alpha_{eh} m_{e_i}$ \\ \\
$\overline{u}_i u_i h$ & & $-\frac{ig}{2 M_W}\frac{\cos \alpha}{\sin \beta} m_{u_i}$ \\ \\
$\overline{d}_i d_i h$ & & $\frac{ig}{2 M_W}\alpha_{dh} m_{d_i}$ \\ \\
$\overline{e}_i e_i H$ & & $\frac{ig}{2 M_W}\alpha_{eH} m_{e_i}$ \\ \\
$\overline{u}_i u_i H$ & & $-\frac{ig}{2 M_W}\frac{\sin \alpha}{\sin \beta} m_{u_i}$ \\ \\
$\overline{d}_i d_i H$ & & $\frac{ig}{2 M_W}\alpha_{dH} m_{d_i}$ \\ \\
$\overline{e}_i e_i A$ & & $-\frac{g}{2 M_W}\beta_{e} m_{e_i} \gamma_5$ \\ \\
$\overline{u}_i u_i A$ & & $-\frac{g}{2 M_W} \cot \beta m_{u_i} \gamma_5$ \\ \\
$\overline{d}_i d_i A$ & & $-\frac{g}{2 M_W}\beta_{d} m_{d_i} \gamma_5$ \\ \\
$\overline{e}_i e_i G_0$ & & $\frac{g}{2 M_W} m_{e_i} \gamma_5$ \\ \\
\end{tabular}

\newpage
\begin{tabular}{lcl}
$\overline{u}_i u_i G_0$ & & $-\frac{g}{2 M_W} m_{u_i} \gamma_5$ \\ \\
$\overline{d}_i d_i G_0$ & & $\frac{g}{2 M_W} m_{d_i} \gamma_5$ \\ \\
$\overline{e}_i \nu_i H^+$ & & $\frac{ig}{2 \sqrt{2}M_W}\beta_{e} m_{e_i} (1 + \gamma_5)$ \\ \\
$\overline{u}_i d_j H^+$ & & $\frac{ig}{2 \sqrt{2}M_W} V_{ij} \left[ \beta_{d} m_{d_j} (1 + \gamma_5) + \cot \beta m_{u_i} (1 - \gamma_5) \right]$ \\ \\
$\overline{\nu}_i e_i H^-$ & & $\frac{ig}{2 \sqrt{2}M_W}\beta_{e} m_{e_i} (1 - \gamma_5)$ \\ \\
$\overline{d}_i u_j H^-$ & & $\frac{ig}{2 \sqrt{2}M_W} V_{ij}^* \left[ \beta_{d} m_{d_i} (1 - \gamma_5) + \cot \beta m_{u_j} (1 + \gamma_5) \right]$ \\ \\
$\overline{e}_i \nu_i G^+$ & & $-\frac{ig}{2 \sqrt{2}M_W} m_{e_i} (1 + \gamma_5)$ \\ \\
$\overline{u}_i d_j G^+$ & & $\frac{ig}{2 \sqrt{2}M_W} V_{ij} \left[ -m_{d_j} (1 + \gamma_5) + m_{u_i} (1 - \gamma_5) \right]$ \\ \\
$\overline{\nu}_i e_i G^-$ & & $-\frac{ig}{2 \sqrt{2}M_W} m_{e_i} (1 - \gamma_5)$ \\ \\
$\overline{d}_i u_j G^-$ & & $\frac{ig}{2 \sqrt{2}M_W}V_{ij}^* \left[ -m_{d_i} (1 - \gamma_5) + m_{u_j} (1 + \gamma_5) \right]$ \\ \\
\end{tabular}

\newpage
\subsection{Gauge boson-scalar vertices}

\begin{tabular}{lcl}
$h Z_{\mu} Z_{\nu}$ & & $-igM_W \sin \delta g_{\mu \nu}$ \\ \\
$h W_{\mu}^+ W_{\nu}^-$ & & $-igM_W \sin \delta g_{\mu \nu}$ \\ \\
$H Z_{\mu} Z_{\nu}$ & & $igM_W \cos \delta g_{\mu \nu}$ \\ \\
$H W_{\mu}^+ W_{\nu}^-$ & & $igM_W \cos \delta g_{\mu \nu}$ \\ \\
$h h Z_{\mu} Z_{\nu}$ & & $\frac{ig^2}{2} g_{\mu \nu}$ \\ \\
$h h W_{\mu}^+ W_{\nu}^-$ & & $\frac{ig^2}{2} g_{\mu \nu}$ \\ \\
$H H Z_{\mu} Z_{\nu}$ & & $\frac{ig^2}{2}g_{\mu \nu}$ \\ \\
$H H W_{\mu}^+ W_{\nu}^-$ & & $\frac{ig^2}{2} g_{\mu \nu}$ \\ \\
$A A Z_{\mu} Z_{\nu}$ & & $\frac{ig^2}{2} g_{\mu \nu}$ \\ \\
$A A W_{\mu}^+ W_{\nu}^-$ & & $\frac{ig^2}{2} g_{\mu \nu}$ \\ \\
$G_0 G_0 Z_{\mu} Z_{\nu}$ & & $\frac{ig^2}{2}g_{\mu \nu}$ \\ \\
\end{tabular}

\newpage
\begin{tabular}{lcl}
$G_0 G_0 W_{\mu}^+ W_{\nu}^-$ & & $\frac{ig^2}{2} g_{\mu \nu}$ \\ \\
$H^+ H^- Z_{\mu} Z_{\nu}$ & & $2ie^2 \cot^2(2 \theta_W) g_{\mu \nu}$ \\ \\
$H^+ H^- W_{\mu}^+ W_{\nu}^-$ & & $\frac{ig^2}{2} g_{\mu \nu}$ \\ \\
$G^+ G^- Z_{\mu} Z_{\nu}$ & & $2ie^2 \cot^2(2 \theta_W) g_{\mu \nu}$ \\ \\
$G^+ G^- W_{\mu}^+ W_{\nu}^-$ & & $\frac{ig^2}{2} g_{\mu \nu}$ \\ \\
$H^+ H^- A_{\mu} A_{\nu}$ & & $2ie^2 g_{\mu \nu}$ \\ \\
$G^+ G^- A_{\mu} A_{\nu}$ & & $2ie^2 g_{\mu \nu}$ \\ \\
$H^+ H^- A_{\mu} Z_{\nu}$ & & $-2ie^2 \cot^2(2 \theta_W) g_{\mu \nu}$ \\ \\
$G^+ G^- A_{\mu} Z_{\nu}$ & & $-2ie^2 \cot^2(2 \theta_W) g_{\mu \nu}$ \\ \\
$H^{\pm} A W_{\mu}^{\mp} A_{\nu}$ & & $\pm \frac{g^2 \sin (\theta_W)}{2}g_{\mu \nu}$ \\ \\
\end{tabular}

\newpage

\begin{tabular}{lcl}
$H^{\pm} A W_{\mu}^{\mp} Z_{\nu}$ & & $\pm \frac{g^2 \sin (\theta_W)}{2} \tan (\theta_W) g_{\mu \nu}$ \\ \\
$G^{\pm} G_0 W_{\mu}^{\mp} A_{\nu}$ & & $\pm \frac{g^2 \sin (\theta_W)}{2} g_{\mu \nu}$ \\ \\
$G^{\pm} G_0 W_{\mu}^{\mp} Z_{\nu}$ & & $\pm \frac{g^2 \sin (\theta_W)}{2} \tan (\theta_W) g_{\mu \nu}$ \\ \\
$H^{\pm} h W_{\mu}^{\mp} A_{\nu}$ & & $\cos \delta \frac{ig^2 \sin (\theta_W)}{2} g_{\mu \nu}$ \\ \\
$H^{\pm} h W_{\mu}^{\mp} Z_{\nu}$ & & $\cos\delta \frac{ig^2 \sin (\theta_W)}{2} \tan (\theta_W) g_{\mu \nu}$ \\ \\
$G^{\pm} H W_{\mu}^{\mp} A_{\nu}$ & & $\cos \delta \frac{ig^2 \sin (\theta_W)}{2} g_{\mu \nu}$ \\ \\
$G^{\pm} H W_{\mu}^{\mp} Z_{\nu}$ & & $\cos\delta \frac{ig^2 \sin (\theta_W)}{2} \tan (\theta_W) g_{\mu \nu}$ \\ \\
$H^{\pm} H W_{\mu}^{\mp} A_{\nu}$ & & $\sin \delta \frac{ig^2 \sin (\theta_W)}{2} g_{\mu \nu}$ \\ \\
$H^{\pm} H W_{\mu}^{\mp} Z_{\nu}$ & & $\sin \delta \frac{ig^2 \sin (\theta_W)}{2} \tan (\theta_W) g_{\mu \nu}$ \\ \\
\end{tabular}

\newpage

\begin{tabular}{lcl}
$G^{\pm} h W_{\mu}^{\mp} A_{\nu}$ & & $- \sin \delta \frac{ig^2 \sin (\theta_W)}{2} g_{\mu \nu}$ \\ \\
$G^{\pm} h W_{\mu}^{\mp} Z_{\nu}$ & & $- \sin \delta \frac{ig^2 \sin (\theta_W)}{2}\tan (\theta_W) g_{\mu \nu}$ \\ \\
$G^{\pm}W_{\mu}^{\mp} A_{\nu}$ & & $i M_W e g_{\mu \nu}$ \\ \\
$G^{\pm}W_{\mu}^{\mp} Z_{\nu}$ & & $i M_W e \tan \theta_W g_{\mu \nu}$ \\ \\
$A_{\mu} H^+ H^-$ & & $i e (p_{H^+} - p_{H^-})_{\mu}$ \\ \\
$Z_{\mu} H^+ H^-$ & & $- i e \cot (2 \theta_W)(p_{H^+} - p_{H^-})_{\mu}$ \\ \\
$A_{\mu} G^+ G^-$ & & $i e (p_{G^+} - p_{G^-})_{\mu}$ \\ \\
$Z_{\mu} G^+ G^-$ & & $- i e \cot (2 \theta_W)(p_{G^+} - p_{G^-})_{\mu}$ \\ \\
$Z_{\mu} H G_0$ & & $\frac{gM_Z}{2M_W} \cos \delta (p_{H} - p_{G_0})_{\mu}$ \\ \\
$Z_{\mu} h A$ & & $\frac{gM_Z}{2M_W} \cos \delta (p_{h} - p_{A})_{\mu}$ \\ \\
$Z_{\mu} H A$ & & $\frac{gM_Z}{2M_W} \sin \delta (p_{H} - p_{A})_{\mu}$ \\ \\
\end{tabular}

\newpage

\begin{tabular}{lcl}
$Z_{\mu} h G_0$ & & $\frac{gM_Z}{2M_W} \sin \delta (p_{G_0} - p_{h})_{\mu}$ \\ \\
$W_{\mu}^{\mp} H^{\pm}A$ & & $\frac{g}{2}(p_{H^{\pm}} - p_{A})_{\mu}$ \\ \\
$W_{\mu}^{\mp}G^{\pm}G_0$ & & $\frac{g}{2}(p_{G^{\pm}} - p_{G_0})_{\mu}$ \\ \\
$W_{\mu}^{\mp}H^{\pm}h$ & & $\pm i \cos \delta \frac{g}{2}(p_{H^{\pm}} - p_{h})_{\mu}$ \\ \\
$W_{\mu}^{\mp}G^{\pm}H$ & & $\pm i \cos \delta \frac{g}{2}(p_{G^{\pm}} - p_{H})_{\mu}$ \\ \\
$W_{\mu}^{\mp}H^{\pm}H$ & & $\pm i \sin \delta \frac{g}{2}(p_{H^{\pm}} - p_{H})_{\mu}$ \\ \\
$W_{\mu}^{\mp}G^{\pm}h$ & & $\mp i \sin \delta \frac{g}{2}(p_{G^{\pm}} - p_{h})_{\mu}$ \\ \\
\end{tabular}

\newpage
\subsection{Ghost-scalar vertices}

\begin{tabular}{lcl}
$\overline{C}^{\pm} C^{\mp}H$ & & $-\frac{igM_W}{2} \cos \delta$ \\ \\
$\overline{C}^{\pm} C^{\mp}h$ & & $\frac{igM_W}{2} \sin \delta$ \\ \\
$\overline{C}^{\pm} C^{\mp}G_0$ & & $\pm \frac{igM_W}{2}$ \\ \\
$\overline{C}_{Z} C_{Z}H$ & & $- \frac{igM_Z}{2 \cos \theta_W} \cos \delta$ \\ \\
$\overline{C}_{Z} C_{Z}h$ & & $ \frac{igM_Z}{2 \cos \theta_W} \sin \delta$ \\ \\
$\overline{C}^{\pm} C_Z G^{\mp}$ & & $\pm \frac{igM_Z}{2}\cos (2 \theta_W)$ \\ \\
$\overline{C}^{\pm} C_A G^{\mp}$ & & $\mp ieM_W$ \\ \\
$\overline{C}_Z C^{\pm} G^{\mp}$ & & $-\frac{igM_Z}{2}$ \\ \\
\end{tabular}

\newpage

\section{Renormalization constants}
\subsection{Two-point functions}
\subsubsection{Scalar counterterms}

In the CP-even scalar sector the six renormalization constants, $Z_{HH}$,  $Z_{hh}$, $Z_{Hh}$, $Z_{hH}$, $\delta M_{H}^2$ and \(\delta M_{h}^2\) are determined by solving the following set of equations: 

\setcounter{equation}{0}
\renewcommand{\theequation}{\mbox{B -} \arabic{equation}}
\begin{eqnarray}
& &  \Sigma_{H H} ( M_{H}^2) - Z_{H H} \delta M_{H}^2 - Z_{h H} (M_{h}^2 - M_{H}^2 + \delta M_{h}^2) - 2 \frac{T_{\alpha \beta}+\sin^2 \alpha T_{\delta}}{v \sin 2 \beta} = 0 \\
& & \frac{d}{dq^2} \Sigma_{H H} ( M_{H}^2) + Z_{H H} + Z_{hH} = 0
\end{eqnarray}

\begin{eqnarray}
& & \Sigma_{h h} ( M_{h}^2) - Z_{h h} \delta M_{h}^2 - Z_{H h} (M_{H}^2 - M_{h}^2 + \delta M_{H}^2) - 2 \frac{T_{\alpha \beta}+\cos^2 \alpha T_{\delta}}{v \sin 2 \beta} = 0 \\
& &\frac{d}{dq^2} \Sigma_{h h} ( M_{h}^2) + Z_{h h} + Z_{Hh} = 0 
\end{eqnarray}

\begin{eqnarray}
& & \Sigma_{H h} ( M_{H}^2) - Z_{H H}^{1/2} Z_{H h}^{1/2} \delta M_{H}^2 - Z_{h h}^{1/2}Z_{h H}^{1/2} (M_{h}^2 - M_{H}^2 + \delta M_{h}^2) - 2 \frac{\sin 2 \alpha T_{\delta}}{v \sin 2 \beta} = 0  \\
& & \Sigma_{H h} ( M_{h}^2) - Z_{h h}^{1/2} Z_{h H}^{1/2} \delta M_{h}^2 - Z_{H H}^{1/2}Z_{H h}^{1/2} (M_{H}^2 - M_{h}^2 + \delta M_{H}^2) - 2 \frac{\sin 2 \alpha T_{\delta}}{v \sin 2 \beta} = 0 \enskip .
\end{eqnarray}

The CP-odd scalar sector has five renormalization constants to be determined, $Z_{AA}$, $Z_{G_0 G_0}$, $Z_{AG_0}$, $Z_{G_0 A}$ and \(\delta M_{A}^2\) , because the Goldstone boson $G_0$ is massless. From the following set of 6 equations only five are independent due to the Ward identity equivalent to eq. (40) but for the neutral sector: 

\begin{eqnarray}
& &  \Sigma_{AA} ( M_{A}^2) - Z_{AA} \delta M_{A}^2 - Z_{G_0 A}M_{A}^2 - 2 \frac{T_{\alpha \beta}+\cos^2 \beta T_{\delta}}{v \sin 2 \beta} = 0  \\
& & \frac{d}{dq^2} \Sigma_{AA} ( M_{A}^2) + Z_{AA} + Z_{G_0 A} = 0
\end{eqnarray}

\begin{eqnarray}
& & \Sigma_{G_0 G_0} (0) - Z_{A G_0}(M_{A}^2 + \delta M_{A}^2)- 2 \frac{T_{\alpha \beta}+\sin^2 \beta T_{\delta}}{v \sin 2 \beta} = 0 \\
& &\frac{d}{dq^2} \Sigma_{G_0 G_0} ( 0 ) + Z_{G_0 G_0} + Z_{A G_0} = 0 
\end{eqnarray}

\begin{eqnarray}
& & \Sigma_{A G_0} ( M_{A}^2) - Z_{AA}^{1/2} Z_{A G_0}^{1/2} \delta M_{A}^2 + Z_{G_0 G_0}^{1/2}Z_{G_0 A}^{1/2} M_{A}^2 -  \frac{T_{\delta}}{v} = 0 \\
& & \Sigma_{A G_0} (0) - Z_{AA}^{1/2} Z_{A G_0}^{1/2}(M_{A}^2 + \delta M_{A}^2) - \frac{T_{\delta}}{v} = 0 \enskip .  
\end{eqnarray}

Finally, the charged sector behaves like the CP-even one. The five renormalization constants to be determined are, in this case, $Z_{H^+ H^+}$, $Z_{G^+ G^+}$, $Z_{H^+G^+}$, $Z_{G^+ H^+}$ and \(\delta M_{H^+}^2\). The equations are: 

\begin{eqnarray}
& &  \Sigma_{H^+H^+} ( M_{H^+}^2) - Z_{H^+H^+} \delta M_{H^+}^2 - Z_{G^+ H^+}M_{H^+}^2 - 2 \frac{T_{\alpha \beta}+\cos^2 \beta T_{\delta}}{v \sin 2 \beta} = 0  \\
& & \frac{d}{dq^2} \Sigma_{H^+H^+} ( M_{H^+}^2) + Z_{H^+H^+} + Z_{G^+ H^+} = 0
\end{eqnarray}

\begin{eqnarray}
& & \Sigma_{G^+ G^+} (0) - Z_{H^+ G^+}(M_{H^+}^2 + \delta M_{H^+}^2)- 2 \frac{T_{\alpha \beta}+\sin^2 \beta T_{\delta}}{v \sin 2 \beta} = 0  \\
& &\frac{d}{dq^2} \Sigma_{G^+ G^+} ( 0 ) + Z_{G^+ G^+} + Z_{H^+ G^+} = 0 
\end{eqnarray}

\begin{eqnarray}
& & \Sigma_{H^+ G^+} ( M_{H^+}^2) - Z_{H^+H^+}^{1/2} Z_{H^+ G^+}^{1/2} \delta M_{H^+}^2 + Z_{G^+ G^+}^{1/2}Z_{G^+ H^+}^{1/2} M_{H^+}^2 - \frac{T_{\delta}}{v} = 0  \\
& & \Sigma_{H^+ G^+} (0) - Z_{H^+H^+}^{1/2} Z_{H^+ G^+}^{1/2}(M_{H^+}^2 + \delta M_{H^+}^2) - \frac{T_{\delta}}{v} = 0 \enskip . 
\end{eqnarray}
The quantities $T_{\delta}$ and $T_{\alpha \beta}$ are defined in equations (11).

\subsubsection{Mixed counterterms}

The complete set of counterterms for the mixed gauge-scalar sector can be written as

\begin{equation}
Z_{W^+ G^+}^{1/2}  = i k_{\mu}(M_{W}^2 + \delta M_{W}^2)^{1/2} Z_W^{1/2}Z_{G^+ G^+}^{1/2}
\end{equation}
\begin{equation}
Z_{W^+ H^+}^{1/2}  = i k_{\mu}(M_{W}^2 + \delta M_{W}^2)^{1/2} Z_W^{1/2}Z_{G^+ H^+}^{1/2}
\end{equation}

\begin{equation}
Z_{Z G_0}^{1/2}  = i k_{\mu}(M_{Z}^2 + \delta M_{Z}^2)^{1/2} Z_{ZZ}^{1/2}Z_{G_0 G_0}^{1/2}
\end{equation}
\begin{equation}
Z_{Z A}^{1/2}  = i k_{\mu}(M_{Z}^2 + \delta M_{Z}^2)^{1/2} Z_{ZZ}^{1/2}Z_{G_0 A}^{1/2}
\end{equation}

\begin{equation}
Z_{\gamma G_0}^{1/2}  = i k_{\mu}(M_{Z}^2 + \delta M_{Z}^2)^{1/2} Z_{Z \gamma}^{1/2}Z_{G_0 G_0}^{1/2}
\end{equation}
\begin{equation}
Z_{\gamma A}^{1/2}  = i k_{\mu}(M_{Z}^2 + \delta M_{Z}^2)^{1/2} Z_{Z \gamma}^{1/2}Z_{G_0 A}^{1/2} \enskip .
\end{equation}

\subsection{Three and four-point functions}

In this section we present the counterterms for the three and four-point functions involving scalar and other fields. The scalar-scalar couterterms will not be shown since Higgs scattering and Higgs decay involving scalar particles only, in both inicial and final states, is already calculated at tree level and was never observed experimentally. So, there is no point in doing loop corrections to processes not yet observed. However it is staightforward to deduce any of those counterterms: first rewritte the scalar lagrangean as a function of the renormalized fields; then group all terms with the same number and type of fields; the factor that multipies those fields is the field renormalization factor; finally renormalize the coupling. 

All along this section we concentrate on the field renormalization. We use $\gamma_{\mu}$ instead of $A_{\mu}$ to represent the photon field so that it will not be confused with the pseudo-scalar field A. The parameter renormalization is written simbolically as $\delta g_{ijk}$ and $\delta g_{ijkl}$ where i,j,k and l are the fields in the vertex. These quantities are determined by a simple variation of the independent parameters in the vertex. We have chosen as free parameters the particle masses, the electric charge, the two angles in the scalar sector ($\alpha$ and $\beta$) and the four independent angles in the CKM matrix. We use several constants as a bookeeping to make the vertex expressions simpler. Among them are $g$, the SU(2) gauge constant, $\theta_{W}$, the Weinberg angle, the angle $\delta=\alpha - \beta$ and the couplings expressed in table 2. The first three can be written in terms of the independent parameters as:

\begin{equation} 
\frac{\delta g}{g} =\frac{ \delta e}{e}+\frac{M_W^2}{2(M_Z^2-M_W^2)}\left[ \frac{\delta M_W^2}{M_W^2}-  \frac{\delta M_Z^2}{M_Z^2} \right] 
\end{equation}

\begin{equation} 
\delta \theta_W =- \frac{\delta M_W}{(M_Z^2-M_W^2)^{1/2}}+ \frac{M_W \delta M_Z}{M_Z (M_Z^2-M_W^2)^{1/2}} 
\end{equation}

\begin{equation} 
\delta (\delta) =\delta \alpha - \delta \beta \enskip . 
\end{equation}

The parameter renormalization in the vertices is easily calculated and so we will just give an example of how it is done. In the example we will use the vertex $g_{\overline{e}_i e_i h}$

\begin{equation} 
\delta g_{\overline{e}_i e_i h} =\delta g \frac{i \alpha_{eh} m_{e_i}}{2 M_W}- \delta M_W 
\frac{i g \alpha_{eh} m_{e_i}}{2 M_W^2}+ (\delta  \alpha_{eh} m_{e_i} + \delta m_{e_i} \alpha_{eh}) \frac{i g}{2 M_W}
\end{equation}
with
\begin{equation}
\delta  \alpha_{eh}=\frac{\sin \alpha}{\sin \beta}\delta \alpha - \frac{\cos \alpha \cos \beta}{\sin^2 \beta}\delta \beta \enskip. \nonumber
\end{equation}

\subsubsection{1 scalar + 2 gauge}
\begin{tabular}{lcl}
$h Z_{\mu} Z_{\nu}$ & & $g_{hZZ} Z_{hh}^{1/2}Z_{ZZ}+ g_{HZZ} Z_{Hh}^{1/2}Z_{ZZ} + \delta g_{hZZ} $ \\ \\
$H Z_{\mu} Z_{\nu}$ & & $g_{HZZ} Z_{HH}^{1/2}Z_{ZZ}+ g_{hZZ} Z_{hH}^{1/2}Z_{ZZ} + \delta g_{HZZ} $ \\ \\
$h \gamma_{\mu} \gamma_{\nu}$ & & $g_{hZZ} Z_{hh}^{1/2}Z_{Z \gamma}+ g_{HZZ} Z_{Hh}^{1/2}Z_{Z \gamma}$ \\ \\
$H \gamma_{\mu} \gamma_{\nu}$ & & $g_{HZZ} Z_{HH}^{1/2}Z_{Z \gamma}+ g_{hZZ} Z_{hH}^{1/2}Z_{Z \gamma}$ \\ \\
$h Z_{\mu} \gamma_{\nu}$ & & $2g_{hZZ} Z_{hh}^{1/2} Z_{ZZ}^{1/2} Z_{Z \gamma}^{1/2} + 2g_{HZZ} Z_{Hh}^{1/2}Z_{ZZ}^{1/2} Z_{Z \gamma}^{1/2}$ \\ \\
\end{tabular}

\newpage
\begin{tabular}{lcl}
$H Z_{\mu} \gamma_{\nu}$ & & $2g_{HZZ} Z_{HH}^{1/2} Z_{ZZ}^{1/2} Z_{Z \gamma}^{1/2} + 2g_{hZZ} Z_{hH}^{1/2}Z_{ZZ}^{1/2} Z_{Z \gamma}^{1/2}$ \\ \\
$h W_{\mu}^+ W_{\nu}^-$ & & $g_{hWW} Z_{hh}^{1/2} Z_{W} + g_{HWW} Z_{Hh}^{1/2}Z_{W}+ \delta g_{hWW}$ \\ \\
$H W_{\mu}^+ W_{\nu}^-$ & & $g_{HWW} Z_{HH}^{1/2} Z_{W} + g_{hWW} Z_{hH}^{1/2}Z_{W}+ \delta g_{HWW}$ \\ \\
$G^{\pm} W_{\mu}^{\mp} \gamma_{\nu}$ & & $g_{G^+ W \gamma} Z_{G^+ G^+}^{1/2} Z_{W}^{1/2} Z_{\gamma \gamma}^{1/2} + g_{G^+ W Z} Z_{G^+ G^+}^{1/2} Z_{W}^{1/2} Z_{Z \gamma}^{1/2} + \delta g_{G^+ W \gamma}$ \\ \\
$G^{\pm} W_{\mu}^{\mp} Z_{\nu}$ & & $g_{G^+ W Z} Z_{G^+ G^+}^{1/2} Z_{W}^{1/2} Z_{ZZ}^{1/2} + g_{G^+ W \gamma} Z_{G^+ G^+}^{1/2} Z_{W}^{1/2} Z_{ \gamma Z}^{1/2} + \delta g_{G^+ W Z}$ \\ \\
$H^{\pm} W_{\mu}^{\mp} \gamma_{\nu}$ & & $g_{G^+ W \gamma} Z_{G^+ H^+}^{1/2} Z_{W}^{1/2} Z_{\gamma \gamma}^{1/2} + g_{G^+ W Z} Z_{G^+ H^+}^{1/2} Z_{W}^{1/2} Z_{Z \gamma}^{1/2}$ \\ \\
$H^{\pm} W_{\mu}^{\mp} Z_{\nu}$ & & $g_{G^+ W Z} Z_{G^+ H^+}^{1/2} Z_{W}^{1/2} Z_{ZZ}^{1/2} + g_{G^+ W \gamma} Z_{G^+ H^+}^{1/2} Z_{W}^{1/2} Z_{ \gamma Z}^{1/2} $ \\ \\
\end{tabular}

\subsubsection{2 scalar + 2 gauge}
\begin{tabular}{lcl}
$hh Z_{\mu} Z_{\nu}$ & & $g_{hhZZ} Z_{hh}Z_{ZZ}+ g_{HHZZ} Z_{Hh}Z_{ZZ} + \delta g_{hhZZ} $ \\ \\
$HH Z_{\mu} Z_{\nu}$ & & $g_{HHZZ} Z_{HH}Z_{ZZ}+ g_{hhZZ} Z_{hH}Z_{ZZ} + \delta g_{HHZZ} $ \\ \\
$hh \gamma_{\mu} \gamma_{\nu}$ & & $g_{hhZZ} Z_{hh}Z_{Z \gamma}+ g_{HHZZ} Z_{Hh}Z_{Z \gamma}$ \\ \\
\end{tabular}

\newpage
\begin{tabular}{lcl}
$HH \gamma_{\mu} \gamma_{\nu}$ & & $g_{HHZZ} Z_{HH}Z_{Z \gamma}+ g_{hhZZ} Z_{hH}Z_{Z \gamma}$ \\ \\
$hh Z_{\mu} \gamma_{\nu}$ & & $2g_{hhZZ} Z_{hh}Z_{ZZ}^{1/2} Z_{Z \gamma}^{1/2}+ 2g_{HHZZ} Z_{Hh}Z_{ZZ}^{1/2} Z_{Z \gamma}^{1/2}$ \\ \\
$HH Z_{\mu} \gamma_{\nu}$ & & $2g_{HHZZ} Z_{HH}Z_{ZZ}^{1/2} Z_{Z \gamma}^{1/2}+ 2g_{hhZZ} Z_{hH}Z_{ZZ}^{1/2} Z_{Z \gamma}^{1/2}$ \\ \\
$hH Z_{\mu} Z_{\nu}$ & & $2g_{HHZZ} Z_{HH}^{1/2} Z_{Hh}^{1/2} Z_{ZZ}+ 2g_{hhZZ} Z_{hh}^{1/2} Z_{hH}^{1/2} Z_{Z Z}$ \\ \\
$hH \gamma_{\mu} \gamma_{\nu}$ & & $2g_{hhZZ} Z_{hh}^{1/2} Z_{hH}^{1/2} Z_{Z \gamma}+ 2g_{HHZZ} Z_{HH}^{1/2} Z_{Hh}^{1/2} Z_{Z \gamma}$ \\ \\
$hH Z_{\mu} \gamma_{\nu}$ & & $4g_{hhZZ} Z_{hh}^{1/2} Z_{hH}^{1/2} Z_{ZZ}^{1/2} Z_{Z \gamma}^{1/2}+ 4g_{HHZZ} Z_{HH}^{1/2} Z_{Hh}^{1/2} Z_{ZZ}^{1/2} Z_{Z \gamma}^{1/2}$ \\ \\
$hh W_{\mu}^+ W_{\nu}^-$ & & $g_{hhWW} Z_{hh}Z_{W}+ g_{HHWW} Z_{Hh}Z_{W} + \delta g_{hhWW} $ \\ \\
$HH W_{\mu}^+ W_{\nu}^-$ & & $g_{HHWW} Z_{HH}Z_{W}+ g_{hhWW} Z_{hH}Z_{W} + \delta g_{HHWW} $ \\ \\
$hH W_{\mu}^+ W_{\nu}^-$ & & $2g_{hhWW} Z_{hh}^{1/2} Z_{hH}^{1/2} Z_{W}+ 2g_{HHWW} Z_{HH}^{1/2} Z_{Hh}^{1/2} Z_{W}$ \\ \\
$AA Z_{\mu} Z_{\nu}$ & & $g_{AAZZ} Z_{AA}Z_{ZZ}+ g_{G_0 G_0 ZZ} Z_{G_0 A}Z_{ZZ} + \delta g_{AAZZ} $ \\ \\
$G_0 G_0 Z_{\mu} Z_{\nu}$ & & $ g_{G_0 G_0 ZZ}  Z_{G_0 G_0}Z_{ZZ}+ g_{AAZZ}  Z_{A G_0}Z_{ZZ}+ \delta g_{G_0 G_0 ZZ} $ \\ \\
\end{tabular}

\newpage
\begin{tabular}{lcl}
$AA \gamma_{\mu} \gamma_{\nu}$ & & $g_{AAZZ} Z_{AA}Z_{Z \gamma}+ g_{G_0 G_0 ZZ} Z_{G_0 A}Z_{Z \gamma}$ \\ \\
$G_0 G_0  \gamma_{\mu} \gamma_{\nu}$ & & $ g_{G_0 G_0 ZZ} Z_{G_0 G_0} Z_{Z \gamma} +g_{AAZZ} Z_{AG_0}Z_{Z \gamma}$ \\ \\
$AA Z_{\mu} \gamma_{\nu}$ & & $2g_{AAZZ} Z_{AA}Z_{ZZ}^{1/2} Z_{Z \gamma}^{1/2} + 2g_{G_0 G_0 ZZ} Z_{G_0 A}Z_{ZZ}^{1/2} Z_{Z \gamma}^{1/2}$ \\ \\
$G_0 G_0 Z_{\mu} \gamma_{\nu}$ & & $2g_{AAZZ} Z_{AG_0}Z_{ZZ}^{1/2} Z_{Z \gamma}^{1/2} + 2g_{G_0 G_0 ZZ} Z_{G_0 G_0}Z_{ZZ}^{1/2} Z_{Z \gamma}^{1/2}$ \\ \\
$A G_0 Z_{\mu} Z_{\nu}$ & & $2g_{AAZZ}Z_{AA}^{1/2} Z_{A G_0}^{1/2} Z_{ZZ} + 2g_{G_0 G_0 ZZ} Z_{G_0 G_0}^{1/2} Z_{G_0 A}^{1/2} Z_{Z Z}$ \\ \\
$A G_0 \gamma_{\mu} \gamma_{\nu}$ & & $2g_{AAZZ}Z_{AA}^{1/2} Z_{A G_0}^{1/2} Z_{Z \gamma} + 2g_{G_0 G_0 ZZ} Z_{G_0 G_0}^{1/2} Z_{G_0 A}^{1/2} Z_{Z \gamma}$ \\ \\
$A G_0 Z_{\mu} \gamma_{\nu}$ & & $4g_{AAZZ} Z_{AA}^{1/2} Z_{AG_0}^{1/2} Z_{ZZ}^{1/2} Z_{Z \gamma}^{1/2}+ 4g_{G_0 G_0 ZZ} Z_{G_0 G_0}^{1/2} Z_{G_0 A}^{1/2} Z_{ZZ}^{1/2} Z_{Z \gamma}^{1/2}$ \\ \\
$AA W_{\mu}^+ W_{\nu}^-$ & & $g_{AAWW} Z_{AA}Z_{W}+ g_{G_0 G_0 WW} Z_{G_0 A}Z_{W} + \delta g_{AAWW} $ \\ \\
$G_0 G_0 W_{\mu}^+ W_{\nu}^-$ & & $g_{G_0 G_0 WW} Z_{G_0 G_0}Z_{W}+ g_{AAWW} Z_{AG_0}Z_{W} + \delta g_{G_0 G_0 WW} $ \\ \\
$A G_0 W_{\mu}^+ W_{\nu}^-$ & & $2g_{AAWW} Z_{AA}^{1/2} Z_{AG_0}^{1/2} Z_{W}+ 2g_{G_0 G_0 WW} Z_{G_0 G_0}^{1/2} Z_{G_0 A}^{1/2} Z_{W}$ \\ \\
\end{tabular}

\newpage
\begin{tabular}{lcl}
$H^+ H^- Z_{\mu} Z_{\nu}$ & & $g_{H^+H^-ZZ} Z_{H^+H^-} Z_{ZZ}+ g_{G^+G^-ZZ} Z_{G^+H^-} Z_{ZZ}+ g_{H^+H^- \gamma \gamma} Z_{H^+H^-} Z_{\gamma Z}$ \\ & & $+ g_{G^+G^- \gamma \gamma} Z_{G^+H^-} Z_{\gamma Z}+ g_{H^+H^- \gamma Z} Z_{H^+H^-} Z_{\gamma Z}^{1/2} Z_{ZZ}^{1/2}$ \\ & & $+ g_{G^+G^- \gamma Z} Z_{G^+H^-} Z_{\gamma Z}^{1/2} Z_{ZZ}^{1/2}+ \delta g_{H^+H^-ZZ}$ \\ \\
$G^+ G^- Z_{\mu} Z_{\nu}$ & & $g_{G^+G^-ZZ} Z_{G^+G^-} Z_{ZZ}+ g_{H^+H^-ZZ} Z_{H^+G^-} Z_{ZZ}+ g_{H^+H^- \gamma \gamma} Z_{H^+G^-} Z_{\gamma Z}$ \\ & & $+ g_{G^+G^- \gamma \gamma} Z_{G^+G^-} Z_{\gamma Z}+ g_{H^+H^- \gamma Z} Z_{H^+G^-} Z_{\gamma Z}^{1/2} Z_{ZZ}^{1/2}$ \\ & & $+ g_{G^+G^- \gamma Z} Z_{G^+G^-} Z_{\gamma Z}^{1/2} Z_{ZZ}^{1/2}+ \delta g_{G^+G^-ZZ}$ \\ \\
$H^+ H^- \gamma_{\mu} \gamma_{\nu}$ & & $g_{H^+H^- \gamma \gamma} Z_{H^+H^-} Z_{\gamma \gamma}+ g_{H^+H^-ZZ} Z_{H^+H^-} Z_{Z \gamma}+ g_{G^+G^- ZZ} Z_{G^+H^-} Z_{Z \gamma}$ \\ & & $+ g_{G^+G^- \gamma \gamma} Z_{G^+H^-} Z_{\gamma \gamma}+ g_{H^+H^- \gamma Z} Z_{H^+H^-} Z_{\gamma \gamma}^{1/2} Z_{Z \gamma}^{1/2}$ \\ & & $+ g_{G^+G^- \gamma Z} Z_{G^+H^-} Z_{\gamma \gamma}^{1/2} Z_{Z \gamma}^{1/2}+ \delta g_{H^+H^- \gamma \gamma}$ \\ \\
$G^+ G^- \gamma_{\mu} \gamma_{\nu}$ & & $g_{G^+G^- \gamma \gamma} Z_{G^+G^-} Z_{\gamma \gamma}+ g_{H^+H^-ZZ} Z_{H^+G^-} Z_{Z \gamma}+ g_{G^+G^- ZZ} Z_{G^+G^-} Z_{Z \gamma}$ \\ & & $+ g_{H^+H^- \gamma \gamma} Z_{H^+G^-} Z_{\gamma \gamma}+ g_{H^+H^- \gamma Z} Z_{H^+G^-} Z_{\gamma \gamma}^{1/2} Z_{Z \gamma}^{1/2}$ \\ & & $+ g_{G^+G^- \gamma Z} Z_{G^+G^-} Z_{\gamma \gamma}^{1/2} Z_{Z \gamma}^{1/2}+ \delta g_{G^+G^- \gamma \gamma}$ \\ \\
$H^+ H^- \gamma_{\mu} Z_{\nu}$ & & $g_{H^+H^- \gamma Z} Z_{H^+H^-} (Z_{\gamma \gamma}^{1/2} Z_{ZZ}^{1/2}+ Z_{\gamma Z}^{1/2} Z_{Z \gamma}^{1/2})+ 2g_{H^+H^- ZZ} Z_{H^+H^-} Z_{ZZ}^{1/2}Z_{Z \gamma}^{1/2}  $ \\ & & $+g_{G^+G^-\gamma Z} Z_{G^+H^-} (Z_{\gamma \gamma}^{1/2} Z_{ZZ}^{1/2}+ Z_{\gamma Z}^{1/2} Z_{Z \gamma}^{1/2})+ 2g_{G^+G^- Z Z} Z_{G^+H^-} Z_{ZZ}^{1/2} Z_{Z \gamma}^{1/2}$ \\ & & $+2g_{G^+G^- \gamma \gamma} Z_{G^+H^-} Z_{\gamma \gamma}^{1/2} Z_{\gamma Z}^{1/2}+ 2g_{H^+H^- \gamma \gamma} Z_{H^+H^-} Z_{\gamma \gamma}^{1/2} Z_{\gamma Z}^{1/2}+ \delta g_{H^+H^- \gamma Z}$ \\ \\
\end{tabular}

\newpage
\begin{tabular}{lcl}
$G^+ G^- \gamma_{\mu} Z_{\nu}$ & & $g_{G^+G^- \gamma Z} Z_{G^+G^-} (Z_{\gamma \gamma}^{1/2} Z_{ZZ}^{1/2}+ Z_{\gamma Z}^{1/2} Z_{Z \gamma}^{1/2})+ 2g_{H^+H^- ZZ} Z_{H^+G^-} Z_{ZZ}^{1/2} Z_{Z \gamma}^{1/2}  $ \\ & & $+g_{H^+H^-\gamma Z} Z_{H^+G^-} (Z_{\gamma \gamma}^{1/2} Z_{ZZ}^{1/2}+ Z_{\gamma Z}^{1/2} Z_{Z \gamma}^{1/2})+ 2g_{G^+G^- Z Z} Z_{G^+G^-} Z_{ZZ}^{1/2} Z_{Z \gamma}^{1/2}$ \\ & & 
$+2g_{H^+H^- \gamma \gamma} Z_{H^+G^-} Z_{\gamma \gamma}^{1/2} Z_{\gamma Z}^{1/2}+ 2g_{G^+G^- \gamma \gamma} Z_{G^+G^-} Z_{\gamma \gamma}^{1/2} Z_{\gamma Z}^{1/2}+ \delta g_{G^+G^- \gamma Z}$ \\ \\

$H^{\pm} G^{\mp} Z_{\mu} Z_{\nu}$ & & $g_{H^+H^- Z Z} Z_{H^+H^-}^{1/2} Z_{H^+ G^-}^{1/2} Z_{ZZ}+ g_{G^+G^- ZZ} Z_{G^+G^-}^{1/2} Z_{G^+ H^-}^{1/2} Z_{Z Z} $ \\ & & $+g_{H^+H^-\gamma \gamma} Z_{H^+H^-}^{1/2}Z_{H^+ G^-}^{1/2} Z_{\gamma Z}+ g_{G^+G^- \gamma \gamma} Z_{G^+G^-}^{1/2} Z_{G^+H^-}^{1/2} Z_{\gamma Z}$ \\ & & $+g_{H^+H^- \gamma Z} Z_{H^+H^-}^{1/2} Z_{H^+ G^-}^{1/2} Z_{\gamma Z}^{1/2} Z_{ZZ}^{1/2}+ g_{G^+G^- \gamma Z} Z_{G^+G^-}^{1/2}Z_{G^+H^-}^{1/2} Z_{\gamma Z}^{1/2} Z_{Z Z}^{1/2}$ \\ \\

$H^{\pm} G^{\mp} \gamma_{\mu} \gamma_{\nu}$ & & $g_{H^+H^- Z Z} Z_{H^+H^-}^{1/2} Z_{H^+ G^-}^{1/2} Z_{Z \gamma}+ g_{G^+G^- ZZ} Z_{G^+G^-}^{1/2} Z_{G^+ H^-}^{1/2} Z_{Z \gamma} $ \\ & & $+g_{H^+H^-\gamma \gamma} Z_{H^+H^-}^{1/2}Z_{H^+ G^-}^{1/2} Z_{\gamma \gamma}+ g_{G^+G^- \gamma \gamma} Z_{G^+G^-}^{1/2} Z_{G^+H^-}^{1/2} Z_{\gamma \gamma}$ \\ & & $+g_{H^+H^- \gamma Z} Z_{H^+H^-}^{1/2} Z_{H^+ G^-}^{1/2} Z_{\gamma \gamma}^{1/2} Z_{Z \gamma}^{1/2}+ g_{G^+G^- \gamma Z} Z_{G^+G^-}^{1/2}Z_{G^+H^-}^{1/2} Z_{\gamma \gamma}^{1/2} Z_{Z \gamma}^{1/2}$ \\ \\

$H^{\pm} G^{\mp} \gamma_{\mu} Z_{\nu}$ & & $2g_{H^+H^- Z Z} Z_{H^+H^-}^{1/2} Z_{H^+ G^-}^{1/2} Z_{ZZ}^{1/2}Z_{Z \gamma}^{1/2}+ 2g_{G^+G^- \gamma \gamma} Z_{G^+G^-}^{1/2}Z_{G^+H^-}^{1/2} Z_{\gamma \gamma}^{1/2} Z_{\gamma Z}^{1/2}$ \\ & & $+2g_{G^+G^-ZZ} Z_{G^+G^-}^{1/2}Z_{G^+ H^-}^{1/2} Z_{ZZ}^{1/2}Z_{Z \gamma}^{1/2}+2g_{H^+H^- \gamma \gamma} Z_{H^+H^-}^{1/2} Z_{H^+ G^-}^{1/2} Z_{\gamma \gamma}^{1/2} Z_{\gamma Z}^{1/2}$ \\  & & $+ g_{G^+G^- \gamma Z} Z_{G^+G^-}^{1/2} Z_{G^+H^-}^{1/2}(Z_{\gamma \gamma}^{1/2} Z_{ZZ}^{1/2}+ Z_{\gamma Z}^{1/2} Z_{Z \gamma}^{1/2})$ \\ & & $+ g_{H^+H^- \gamma Z} Z_{H^+H^-}^{1/2} Z_{H^+ G^-}^{1/2}(Z_{\gamma \gamma}^{1/2} Z_{ZZ}^{1/2}+ Z_{\gamma Z}^{1/2} Z_{Z \gamma}^{1/2}) $ \\ \\
\end{tabular}

\newpage
\begin{tabular}{lcl}
$H^+ H^- W_{\mu}^+ W_{\nu}^-$ & & $g_{H^+H^- WW} Z_{H^+H^-} Z_W+g_{G^+G^- WW} Z_{G^+H^-}Z_W+ \delta g_{H^+H^- WW}$ \\ \\

$G^+ G^- W_{\mu}^+ W_{\nu}^-$ & & $g_{G^+G^- WW} Z_{G^+G^-} Z_W+g_{H^+H^- WW} Z_{H^+G^-}Z_W+ \delta g_{G^+G^- WW}$ \\ \\

$H^{\pm} G^{\mp} W_{\mu}^+ W_{\nu}^-$ & & $g_{H^+H^- WW} Z_{H^+H^-}^{1/2} Z_{H^+G^-}^{1/2}Z_W+g_{G^+G^- WW}Z_{G^+G^-}^{1/2}Z_{G^+H^-}^{1/2}Z_W$\\ \\

$H^{\pm} A W_{\mu}^{\mp} \gamma_{\nu}$ & & $g_{H^+AW \gamma} Z_{H^+H^-}^{1/2} Z_{AA}^{1/2} Z_W^{1/2}Z_{\gamma \gamma}^{1/2} + g_{H^+AW Z} Z_{H^+H^-}^{1/2} Z_{AA}^{1/2} Z_W^{1/2}Z_{Z \gamma}^{1/2}$ \\ & & $+ g_{G^+G_0W \gamma} Z_{G^+H^-}^{1/2} Z_{G_0 A}^{1/2} Z_W^{1/2}Z_{\gamma \gamma}^{1/2} +g_{G^+G_0W Z} Z_{G^+H^-}^{1/2} Z_{G_0 A}^{1/2} Z_W^{1/2}Z_{Z \gamma}^{1/2}$ \\ & & $ + \delta g_{H^+AW \gamma}$ \\ \\

$H^{\pm} A W_{\mu}^{\mp} Z_{\nu}$ & & $g_{H^+AW Z} Z_{H^+H^-}^{1/2} Z_{AA}^{1/2} Z_W^{1/2}Z_{ZZ}^{1/2} + g_{H^+AW \gamma} Z_{H^+H^-}^{1/2} Z_{AA}^{1/2} Z_W^{1/2}Z_{\gamma Z}^{1/2}$ \\ & & $+ g_{G^+G_0W \gamma} Z_{G^+H^-}^{1/2} Z_{G_0 A}^{1/2} Z_W^{1/2}Z_{\gamma Z}^{1/2} +g_{G^+G_0W Z} Z_{G^+H^-}^{1/2} Z_{G_0 A}^{1/2} Z_W^{1/2}Z_{Z Z}^{1/2}$ \\ & & $ + \delta g_{H^+AWZ}$ \\ \\

$G^{\pm} G_0 W_{\mu}^{\mp} \gamma_{\nu}$ & & $g_{G^+G_0W \gamma} Z_{G^+G^-}^{1/2} Z_{G_0 G_0}^{1/2} Z_W^{1/2}Z_{\gamma \gamma}^{1/2} + g_{H^+AW \gamma} Z_{H^+G^-}^{1/2} Z_{AG_0}^{1/2} Z_W^{1/2}Z_{\gamma \gamma}^{1/2}$ \\ & & $+ g_{H^+ AW Z} Z_{H^+G^-}^{1/2} Z_{AG_0}^{1/2} Z_W^{1/2}Z_{Z \gamma}^{1/2} +g_{G^+G_0W Z} Z_{G^+G^-}^{1/2} Z_{G_0 G_0}^{1/2} Z_W^{1/2}Z_{Z \gamma}^{1/2}$ \\ & & $ + \delta g_{G^+ G_0W \gamma}$ \\ \\

$G^{\pm} G_0 W_{\mu}^{\mp} Z_{\nu}$ & & $g_{G^+G_0W Z} Z_{G^+G^-}^{1/2} Z_{G_0G_0}^{1/2} Z_W^{1/2}Z_{ZZ}^{1/2} + g_{H^+AW \gamma} Z_{H^+G^-}^{1/2} Z_{AG_0}^{1/2} Z_W^{1/2}Z_{\gamma Z}^{1/2}$ \\ & & $+ g_{H^+AW Z} Z_{H^+G^-}^{1/2} Z_{AG_0}^{1/2} Z_W^{1/2}Z_{ZZ}^{1/2} +g_{G^+G_0W \gamma} Z_{G^+G^-}^{1/2} Z_{G_0 G_0}^{1/2} Z_W^{1/2}Z_{\gamma Z}^{1/2}$ \\ & & $ + \delta g_{G^+G_0WZ}$ \\ \\

\end{tabular}

\newpage
\begin{tabular}{lcl}
$H^{\pm} G_0 W_{\mu}^{\mp} \gamma_{\nu}$ & & $g_{G^+G_0W \gamma} Z_{G^+G^-}^{1/2} Z_{G_0 G_0}^{1/2} Z_W^{1/2}Z_{\gamma \gamma}^{1/2} + g_{H^+AW \gamma} Z_{H^+G^-}^{1/2} Z_{AG_0}^{1/2} Z_W^{1/2}Z_{\gamma \gamma}^{1/2}$ \\ & & $+ g_{H^+ AW Z} Z_{H^+G^-}^{1/2} Z_{AG_0}^{1/2} Z_W^{1/2}Z_{Z \gamma}^{1/2} +g_{G^+G_0W Z} Z_{G^+G^-}^{1/2} Z_{G_0 G_0}^{1/2} Z_W^{1/2}Z_{Z \gamma}^{1/2}$ \\ \\
$H^{\pm} G_0 W_{\mu}^{\mp} Z_{\nu}$ & & $g_{H^+AW \gamma} Z_{H^+H^-}^{1/2} Z_{A G_0}^{1/2} Z_W^{1/2}Z_{\gamma Z}^{1/2} + g_{H^+AW Z} Z_{H^+H^-}^{1/2} Z_{AG_0}^{1/2} Z_W^{1/2}Z_{ZZ}^{1/2}$ \\ & & $+ g_{G^+ G_0W \gamma} Z_{G^+H^-}^{1/2} Z_{G_0 G_0}^{1/2} Z_W^{1/2}Z_{\gamma Z}^{1/2} +g_{G^+G_0W Z} Z_{G^+H^-}^{1/2} Z_{G_0 G_0}^{1/2} Z_W^{1/2}Z_{Z Z}^{1/2}$ \\ \\

$G^{\pm} A W_{\mu}^{\mp} \gamma_{\nu}$ & & $g_{H^+AW \gamma} Z_{H^+G^-}^{1/2} Z_{A A}^{1/2} Z_W^{1/2}Z_{\gamma \gamma}^{1/2} + g_{H^+AW Z} Z_{H^+G^-}^{1/2} Z_{AA}^{1/2} Z_W^{1/2}Z_{Z \gamma}^{1/2}$ \\ & & $+ g_{G^+ G_0W \gamma} Z_{G^+G^-}^{1/2} Z_{G_0 A}^{1/2} Z_W^{1/2}Z_{\gamma \gamma}^{1/2} +g_{G^+G_0W Z} Z_{G^+G^-}^{1/2} Z_{G_0 A}^{1/2} Z_W^{1/2}Z_{Z \gamma}^{1/2}$ \\ \\

$G^{\pm} A W_{\mu}^{\mp} Z_{\nu}$ & & $g_{H^+AW \gamma} Z_{H^+G^-}^{1/2} Z_{A A}^{1/2} Z_W^{1/2}Z_{\gamma Z}^{1/2} + g_{H^+AW Z} Z_{H^+G^-}^{1/2} Z_{AA}^{1/2} Z_W^{1/2}Z_{Z Z}^{1/2}$ \\ & & $+ g_{G^+ G_0W \gamma} Z_{G^+G^-}^{1/2} Z_{G_0 A}^{1/2} Z_W^{1/2}Z_{\gamma Z}^{1/2} +g_{G^+G_0W Z} Z_{G^+G^-}^{1/2} Z_{G_0 A}^{1/2} Z_W^{1/2}Z_{Z Z}^{1/2}$ \\ \\

$H^{\pm} h W_{\mu}^{\mp} \gamma_{\nu}$ & & $g_{H^+h W \gamma} Z_{H^+H^-}^{1/2} Z_{hh}^{1/2} Z_W^{1/2}Z_{\gamma \gamma}^{1/2}+ g_{G^+h W \gamma}Z_{G^+H^-}^{1/2} Z_{hh}^{1/2} Z_W^{1/2} Z_{\gamma \gamma}^{1/2}$ \\ & & $+g_{G^+h WZ} Z_{G^+H^-}^{1/2}Z_{hh}^{1/2} Z_W^{1/2}Z_{Z \gamma}^{1/2}+g_{G^+H W Z} Z_{G^+H^-}^{1/2} Z_{Hh}^{1/2} Z_W^{1/2} Z_{Z \gamma}^{1/2}$ \\  & & $+ g_{G^+H W \gamma} Z_{G^+H^-}^{1/2} Z_{Hh}^{1/2}Z_W^{1/2}Z_{\gamma\gamma}^{1/2} + g_{H^+H W \gamma} Z_{H^+H^-}^{1/2}Z_{Hh}^{1/2}Z_W^{1/2} Z_{\gamma \gamma}^{1/2}$ \\ & & $+ g_{H^+H W Z} Z_{H^+H^-}^{1/2} Z_{Hh}^{1/2}Z_W^{1/2}Z_{Z\gamma}^{1/2} + g_{H^+h W Z} Z_{H^+H^-}^{1/2} Z_{hh}^{1/2}Z_W^{1/2}Z_{Z \gamma}^{1/2}$ \\ &  & $+ \delta g_{H^+h W \gamma}$ \\ \\
\end{tabular}

\newpage
\begin{tabular}{lcl}
$G^{\pm} h W_{\mu}^{\mp} \gamma_{\nu}$ & & $g_{G^+h W \gamma} Z_{G^+G^-}^{1/2} Z_{hh}^{1/2} Z_W^{1/2}Z_{\gamma \gamma}^{1/2}+ g_{H^+h W\gamma}Z_{H^+G^-}^{1/2} Z_{hh}^{1/2} Z_W^{1/2} Z_{\gamma \gamma}^{1/2}$ \\ & & $+g_{G^+h WZ} Z_{G^+G^-}^{1/2}Z_{hh}^{1/2} Z_W^{1/2}Z_{Z \gamma}^{1/2}+g_{G^+H W Z} Z_{G^+G^-}^{1/2} Z_{Hh}^{1/2} Z_W^{1/2} Z_{Z \gamma}^{1/2}$ \\  & & $+ g_{G^+H W \gamma} Z_{G^+G^-}^{1/2} Z_{Hh}^{1/2}Z_W^{1/2}Z_{\gamma\gamma}^{1/2} + g_{H^+H W \gamma} Z_{H^+G^-}^{1/2}Z_{Hh}^{1/2}Z_W^{1/2} Z_{\gamma \gamma}^{1/2}$ \\ & & $+ g_{H^+H W Z} Z_{H^+G^-}^{1/2} Z_{Hh}^{1/2}Z_W^{1/2}Z_{Z\gamma}^{1/2} + g_{H^+h W Z} Z_{H^+G^-}^{1/2} Z_{hh}^{1/2}Z_W^{1/2}Z_{Z \gamma}^{1/2}$ \\ &  & $+ \delta g_{G^+h W \gamma}$ \\ \\

$G^{\pm} h W_{\mu}^{\mp}Z_{\nu}$ & & $g_{G^+h W Z} Z_{G^+G^-}^{1/2} Z_{hh}^{1/2} Z_W^{1/2}Z_{ZZ}^{1/2}+ g_{H^+h W\gamma}Z_{H^+G^-}^{1/2} Z_{hh}^{1/2} Z_W^{1/2} Z_{\gamma Z}^{1/2}$ \\ & & $+g_{G^+h W \gamma} Z_{G^+G^-}^{1/2}Z_{hh}^{1/2} Z_W^{1/2}Z_{\gamma Z}^{1/2}+g_{G^+H W Z} Z_{G^+G^-}^{1/2} Z_{Hh}^{1/2} Z_W^{1/2} Z_{Z Z}^{1/2}$ \\  & & $+ g_{G^+H W \gamma} Z_{G^+G^-}^{1/2} Z_{Hh}^{1/2}Z_W^{1/2}Z_{\gamma Z}^{1/2} + g_{H^+H WZ} Z_{H^+G^-}^{1/2} Z_{Hh}^{1/2}Z_W^{1/2} Z_{ZZ}^{1/2}$ \\ & & $+ g_{H^+h W Z} Z_{H^+G^-}^{1/2} Z_{hh}^{1/2}Z_W^{1/2}Z_{ZZ}^{1/2} + g_{H^+H W \gamma} Z_{H^+G^-}^{1/2} Z_{Hh}^{1/2}Z_W^{1/2}Z_{\gamma Z}^{1/2}$ \\ &  & $+ \delta g_{G^+h W Z}$ \\ \\

$G^{\pm} H W_{\mu}^{\mp}Z_{\nu}$ & & $g_{G^+H W Z} Z_{G^+G^-}^{1/2} Z_{HH}^{1/2} Z_W^{1/2}Z_{ZZ}^{1/2}+ g_{H^+h W\gamma}Z_{H^+G^-}^{1/2} Z_{hH}^{1/2} Z_W^{1/2} Z_{\gamma Z}^{1/2}$ \\ & & $+g_{G^+h W \gamma} Z_{G^+G^-}^{1/2}Z_{hH}^{1/2} Z_W^{1/2}Z_{\gamma Z}^{1/2}+g_{G^+h W Z} Z_{G^+G^-}^{1/2} Z_{hH}^{1/2} Z_W^{1/2} Z_{Z Z}^{1/2}$ \\  & & $+ g_{G^+H W \gamma} Z_{G^+G^-}^{1/2} Z_{HH}^{1/2}Z_W^{1/2}Z_{\gamma Z}^{1/2} + g_{H^+H W \gamma} Z_{H^+G^-}^{1/2} Z_{HH}^{1/2}Z_W^{1/2} Z_{\gamma Z}^{1/2}$ \\ & & $+ g_{H^+H W Z} Z_{H^+G^-}^{1/2} Z_{HH}^{1/2}Z_W^{1/2}Z_{ZZ}^{1/2} + g_{H^+h W Z} Z_{H^+G^-}^{1/2} Z_{hH}^{1/2}Z_W^{1/2}Z_{ZZ}^{1/2}$ \\ &  & $+ \delta g_{G^+H W Z}$ \\ \\

\end{tabular}

\newpage
\begin{tabular}{lcl}

$G^{\pm} H W_{\mu}^{\mp}\gamma_{\nu}$ & & $g_{G^+H W \gamma} Z_{G^+G^-}^{1/2} Z_{HH}^{1/2} Z_W^{1/2}Z_{\gamma \gamma}^{1/2}+ g_{H^+hW\gamma}Z_{H^+G^-}^{1/2} Z_{hH}^{1/2} Z_W^{1/2} Z_{\gamma \gamma}^{1/2}$ \\ & & $+g_{G^+h W \gamma} Z_{G^+G^-}^{1/2}Z_{hH}^{1/2} Z_W^{1/2}Z_{\gamma \gamma}^{1/2}+g_{G^+h W Z} Z_{G^+G^-}^{1/2} Z_{hH}^{1/2} Z_W^{1/2} Z_{Z \gamma}^{1/2}$ \\  & & $+ g_{G^+H W Z} Z_{G^+G^-}^{1/2} Z_{HH}^{1/2}Z_W^{1/2}Z_{Z \gamma}^{1/2} + g_{H^+H W \gamma} Z_{H^+G^-}^{1/2} Z_{HH}^{1/2}Z_W^{1/2} Z_{\gamma \gamma}^{1/2}$ \\ & & $+ g_{H^+H W Z} Z_{H^+G^-}^{1/2} Z_{HH}^{1/2}Z_W^{1/2}Z_{Z \gamma}^{1/2} + g_{H^+h W Z} Z_{H^+G^-}^{1/2} Z_{hH}^{1/2}Z_W^{1/2}Z_{Z \gamma}^{1/2}$ \\ &  & $+ \delta g_{G^+H W \gamma}$ \\ \\

$H^{\pm} H W_{\mu}^{\mp}\gamma_{\nu}$ & & $g_{H^+H W \gamma} Z_{H^+H^-}^{1/2} Z_{HH}^{1/2} Z_W^{1/2}Z_{\gamma \gamma}^{1/2}+ g_{H^+hW\gamma}Z_{H^+H^-}^{1/2} Z_{hH}^{1/2} Z_W^{1/2} Z_{\gamma \gamma}^{1/2}$ \\ & & $+g_{G^+h W \gamma} Z_{G^+H^-}^{1/2}Z_{hH}^{1/2} Z_W^{1/2}Z_{\gamma \gamma}^{1/2}+g_{G^+h W Z} Z_{G^+H^-}^{1/2} Z_{hH}^{1/2} Z_W^{1/2} Z_{Z \gamma}^{1/2}$ \\  & & $+ g_{G^+H W Z} Z_{G^+H^-}^{1/2} Z_{HH}^{1/2}Z_W^{1/2}Z_{Z \gamma}^{1/2} + g_{G^+H W \gamma} Z_{G^+H^-}^{1/2} Z_{HH}^{1/2}Z_W^{1/2} Z_{\gamma \gamma}^{1/2}$ \\ & & $+ g_{H^+H W Z} Z_{H^+H^-}^{1/2} Z_{HH}^{1/2}Z_W^{1/2}Z_{Z \gamma}^{1/2} + g_{H^+h W Z} Z_{H^+H^-}^{1/2} Z_{hH}^{1/2}Z_W^{1/2}Z_{Z \gamma}^{1/2}$ \\ &  & $+ \delta g_{H^+H W \gamma}$ \\ \\

$H^{\pm} H W_{\mu}^{\mp}Z_{\nu}$ & & $g_{H^+H W Z} Z_{H^+H^-}^{1/2} Z_{HH}^{1/2} Z_W^{1/2}Z_{ZZ}^{1/2}+ g_{H^+hW\gamma}Z_{H^+H^-}^{1/2} Z_{hH}^{1/2} Z_W^{1/2} Z_{\gamma Z}^{1/2}$ \\ & & $+g_{G^+h W \gamma} Z_{G^+H^-}^{1/2}Z_{hH}^{1/2} Z_W^{1/2}Z_{\gamma Z}^{1/2}+g_{G^+h W Z} Z_{G^+H^-}^{1/2} Z_{hH}^{1/2} Z_W^{1/2} Z_{Z Z}^{1/2}$ \\  & & $+ g_{G^+H W Z} Z_{G^+H^-}^{1/2} Z_{HH}^{1/2}Z_W^{1/2}Z_{Z Z}^{1/2} + g_{G^+H W \gamma} Z_{G^+H^-}^{1/2} Z_{HH}^{1/2}Z_W^{1/2} Z_{\gamma Z}^{1/2}$ \\ & & $+ g_{H^+H W \gamma} Z_{H^+H^-}^{1/2} Z_{HH}^{1/2}Z_W^{1/2}Z_{\gamma Z}^{1/2} + g_{H^+h W Z} Z_{H^+H^-}^{1/2} Z_{hH}^{1/2}Z_W^{1/2}Z_{Z Z}^{1/2}$ \\ &  & $+ \delta g_{H^+H W Z}$ \\ \\

\end{tabular}

\newpage
\begin{tabular}{lcl}
$H^{\pm} h W_{\mu}^{\mp}Z_{\nu}$ & & $g_{H^+h W Z} Z_{H^+H^-}^{1/2} Z_{hh}^{1/2} Z_W^{1/2}Z_{ZZ}^{1/2}+ g_{H^+hW \gamma}Z_{H^+H^-}^{1/2} Z_{hh}^{1/2} Z_W^{1/2} Z_{\gamma Z}^{1/2}$ \\ & & $+g_{G^+h W \gamma} Z_{G^+H^-}^{1/2}Z_{hh}^{1/2} Z_W^{1/2}Z_{\gamma Z}^{1/2}+g_{G^+h W Z} Z_{G^+H^-}^{1/2} Z_{hh}^{1/2} Z_W^{1/2} Z_{Z Z}^{1/2}$ \\  & & $+ g_{G^+H W Z} Z_{G^+H^-}^{1/2} Z_{Hh}^{1/2}Z_W^{1/2}Z_{Z Z}^{1/2} + g_{G^+H W \gamma} Z_{G^+H^-}^{1/2} Z_{Hh}^{1/2}Z_W^{1/2} Z_{\gamma Z}^{1/2}$ \\ & & $+ g_{H^+H W \gamma} Z_{H^+H^-}^{1/2} Z_{Hh}^{1/2}Z_W^{1/2}Z_{\gamma Z}^{1/2} + g_{H^+H W Z} Z_{H^+H^-}^{1/2} Z_{Hh}^{1/2}Z_W^{1/2}Z_{Z Z}^{1/2}$ \\ &  & $+ \delta g_{H^+h W Z}$
\end{tabular}

\subsubsection{2 scalar + 1 gauge}

\begin{tabular}{lcl}
$H^+ H^- \gamma_{\mu}$ & & $(p_{H^+}-p_{H^-})_{\mu} \left[  g_{H^+H^-\gamma} Z_{H^+H^-} Z_{\gamma \gamma}^{1/2}+ g_{H^+H^- Z}Z_{H^+H^-}Z_{Z \gamma}^{1/2} \right.$ \\ & & $ \left. + g_{G^+ G^- \gamma} Z_{G^+H^-} Z_{\gamma \gamma}^{1/2}+ g_{G^+ G^- Z} Z_{G^+H^-}Z_{Z \gamma}^{1/2}+ \delta g_{H^+H^- \gamma} \right]$ \\ \\

$H^+ H^- Z_{\mu}$ & & $(p_{H^+}-p_{H^-})_{\mu} \left[  g_{H^+H^-Z} Z_{H^+H^-} Z_{ZZ}^{1/2}+ g_{H^+H^- \gamma}Z_{H^+H^-}Z_{\gamma Z}^{1/2} \right.$ \\ & & $ \left. + g_{G^+ G^- \gamma} Z_{G^+H^-} Z_{\gamma Z}^{1/2}+ g_{G^+ G^- Z} Z_{G^+H^-}Z_{Z Z}^{1/2}+ \delta g_{H^+H^- Z} \right]$ \\ \\

$G^+ G^- \gamma_{\mu}$ & & $(p_{G^+}-p_{G^-})_{\mu} \left[  g_{G^+G^-\gamma} Z_{G^+G^-} Z_{\gamma \gamma}^{1/2}+ g_{H^+H^- Z}Z_{H^+G^-}Z_{Z \gamma}^{1/2} \right.$ \\ & & $ \left. + g_{H^+ H^- \gamma} Z_{H^+G^-} Z_{\gamma \gamma}^{1/2}+ g_{G^+ G^- Z} Z_{G^+G^-}Z_{Z \gamma}^{1/2}+ \delta g_{G^+G^- \gamma} \right]$ \\ \\

$G^+ G^- Z_{\mu}$ & & $(p_{G^+}-p_{G^-})_{\mu} \left[  g_{G^+G^-Z} Z_{G^+G^-} Z_{ZZ}^{1/2}+ g_{H^+H^- Z}Z_{H^+G^-}Z_{Z Z}^{1/2} \right.$ \\ & & $ \left. + g_{H^+ H^- \gamma} Z_{H^+G^-} Z_{\gamma Z}^{1/2}+ g_{G^+ G^- \gamma}Z_{G^+G^-} Z_{\gamma Z}^{1/2}+ \delta g_{G^+G^- Z} \right]$
\end{tabular}

\newpage
\begin{tabular}{lcl}
$H^{\pm}G^{\mp} \gamma_{\mu}$ & & $\pm (p_{H^{\pm}}-p_{G^{\mp}})_{\mu} \left[  g_{H^+H^- \gamma} Z_{H^+H^-}^{1/2}Z_{H^+G^-}^{1/2} Z_{\gamma \gamma}^{1/2}+ g_{H^+H^- Z}Z_{H^+H^-}^{1/2} \right.$ \\ & & $ \left. Z_{H^+G^-}^{1/2}Z_{Z \gamma}^{1/2} + g_{G^+ G^- \gamma} Z_{G^+G^-}^{1/2}Z_{G^+H^-}^{1/2} Z_{\gamma\gamma}^{1/2} + g_{G^+ G^- Z}Z_{G^+G^-}^{1/2}Z_{G^+H^-}^{1/2} Z_{Z \gamma}^{1/2}\right]$ \\ \\

$H^{\pm}G^{\mp} Z_{\mu}$ & & $\pm (p_{H^{\pm}}-p_{G^{\mp}})_{\mu} \left[  g_{H^+H^- \gamma} Z_{H^+H^-}^{1/2}Z_{H^+G^-}^{1/2} Z_{\gamma Z}^{1/2}+ g_{H^+H^- Z}Z_{H^+H^-}^{1/2} \right.$ \\ & & $ \left. Z_{H^+G^-}^{1/2}Z_{Z Z}^{1/2} + g_{G^+ G^- \gamma} Z_{G^+G^-}^{1/2}Z_{G^+H^-}^{1/2} Z_{\gamma Z}^{1/2}+ g_{G^+ G^- Z}Z_{G^+G^-}^{1/2}Z_{G^+H^-}^{1/2} Z_{Z Z}^{1/2}\right]$ \\ \\

$H G_0 Z_{\mu}$ & & $(p_H-p_{G_0})_{\mu}Z_{ZZ}^{1/2} \left[  g_{H G_0 Z}Z_{HH}^{1/2} Z_{G_0 G_0}^{1/2} + g_{HAZ}Z_{HH}^{1/2} Z_{AG_0}^{1/2} \right.$ \\ & & $ \left. + g_{hAZ} Z_{hH}^{1/2}Z_{AG_0}^{1/2}+g_{hG_0 Z} Z_{hH}^{1/2}Z_{G_0 G_0}^{1/2}+ \delta  g_{H G_0 Z} \right]$ \\ \\

$H A Z_{\mu}$ & & $(p_H-p_{A})_{\mu}Z_{ZZ}^{1/2} \left[  g_{H A Z}Z_{HH}^{1/2} Z_{AA}^{1/2} + g_{HG_0 Z}Z_{HH}^{1/2} Z_{G_0 A}^{1/2} \right.$ \\ & & $ \left. + g_{hAZ} Z_{hH}^{1/2}Z_{AA}^{1/2}+g_{hG_0 Z} Z_{hH}^{1/2}Z_{G_0 A}^{1/2}+ \delta  g_{HAZ} \right]$ \\ \\

$h A Z_{\mu}$ & & $(p_h-p_{A})_{\mu}Z_{ZZ}^{1/2} \left[  g_{h A Z}Z_{hh}^{1/2} Z_{AA}^{1/2} + g_{HG_0 Z}Z_{Hh}^{1/2} Z_{G_0 A}^{1/2} \right.$ \\ & & $ \left. + g_{HAZ} Z_{Hh}^{1/2}Z_{AA}^{1/2}+g_{hG_0 Z} Z_{hh}^{1/2}Z_{G_0 A}^{1/2}+ \delta  g_{hAZ} \right]$ \\ \\

$h G_0 Z_{\mu}$ & & $(p_h-p_{G_0})_{\mu}Z_{ZZ}^{1/2} \left[  g_{h G_0 Z}Z_{hh}^{1/2} Z_{G_0 G_0}^{1/2} + g_{HG_0 Z}Z_{Hh}^{1/2} Z_{G_0G_0}^{1/2} \right.$ \\ & & $ \left. + g_{HAZ} Z_{Hh}^{1/2}Z_{AG_0}^{1/2}+g_{hA Z} Z_{hh}^{1/2}Z_{AG_0}^{1/2}+ \delta  g_{h G_0 Z} \right]$ \\ \\

$H G_0 \gamma_{\mu}$ & & $(p_H-p_{G_0})_{\mu}Z_{Z \gamma}^{1/2} \left[  g_{H G_0 Z}Z_{HH}^{1/2} Z_{G_0 G_0}^{1/2} + g_{HAZ}Z_{HH}^{1/2} Z_{AG_0}^{1/2} \right.$ \\ & & $ \left. + g_{hAZ} Z_{hH}^{1/2}Z_{AG_0}^{1/2}+g_{hG_0 Z} Z_{hH}^{1/2}Z_{G_0 G_0}^{1/2} \right]$ 
\end{tabular}

\newpage
\begin{tabular}{lcl}

$H A \gamma_{\mu}$ & & $(p_H-p_{A})_{\mu}Z_{Z \gamma}^{1/2} \left[  g_{H A Z}Z_{HH}^{1/2} Z_{AA}^{1/2} + g_{HG_0 Z}Z_{HH}^{1/2} Z_{G_0 A}^{1/2} \right.$ \\ & & $ \left. + g_{hAZ} Z_{hH}^{1/2}Z_{AA}^{1/2}+g_{hG_0 Z} Z_{hH}^{1/2}Z_{G_0 A}^{1/2} \right]$ \\ \\

$h A \gamma_{\mu}$ & & $(p_h-p_{A})_{\mu}Z_{Z \gamma}^{1/2} \left[  g_{h A Z}Z_{hh}^{1/2} Z_{AA}^{1/2} + g_{HG_0 Z}Z_{Hh}^{1/2} Z_{G_0 A}^{1/2} \right.$ \\ & & $ \left. + g_{HAZ} Z_{Hh}^{1/2}Z_{AA}^{1/2}+g_{hG_0 Z} Z_{hh}^{1/2}Z_{G_0 A}^{1/2} \right]$ \\ \\

$h G_0 \gamma_{\mu}$ & & $(p_h-p_{G_0})_{\mu}Z_{Z \gamma}^{1/2} \left[  g_{h G_0 Z}Z_{hh}^{1/2} Z_{G_0 G_0}^{1/2} + g_{HG_0 Z}Z_{Hh}^{1/2} Z_{G_0G_0}^{1/2} \right.$ \\ & & $ \left. + g_{HAZ} Z_{Hh}^{1/2}Z_{AG_0}^{1/2}+g_{hA Z} Z_{hh}^{1/2}Z_{AG_0}^{1/2} \right]$ \\ \\

$H^{\pm}AW_{\mu}^{\mp}$ & & $(p_{H^{\pm}}-p_{A})_{\mu}Z_W^{1/2} \left[  g_{H^+A W}Z_{H^+H^-}^{1/2} Z_{AA}^{1/2} + g_{G^+ G_0W}Z_{G^+H^-}^{1/2} Z_{G_0A}^{1/2} + \delta g_{H^+A W} \right]$ \\ \\

$G^{\pm}G_0W_{\mu}^{\mp}$ & & $(p_{G^{\pm}}-p_{G_0})_{\mu}Z_W^{1/2} \left[  g_{G^+G_0 W}Z_{G^+G^-}^{1/2} Z_{G_0 G_0}^{1/2} + g_{H^+AW} Z_{H^+G^-}^{1/2} Z_{AG_0}^{1/2} + \delta g_{G^+ G_0 W} \right]$ \\ \\

$H^{\pm}G_0W_{\mu}^{\mp}$ & & $(p_{H^{\pm}}-p_{G_0})_{\mu}Z_W^{1/2} \left[  g_{H^+A W}Z_{H^+H^-}^{1/2} Z_{AG_0}^{1/2} + g_{G^+ G_0W}Z_{G^+H^-}^{1/2} Z_{G_0 G_0}^{1/2} \right]$ \\ \\

$G^{\pm}AW_{\mu}^{\mp}$ & & $(p_{G^{\pm}}-p_{A})_{\mu}Z_W^{1/2} \left[  g_{G^+G_0 W}Z_{G^+G^-}^{1/2} Z_{G_0 A}^{1/2} + g_{H^+AW} Z_{H^+G^-}^{1/2} Z_{AA}^{1/2} \right]$ \\ \\

$H^{\pm}h W_{\mu}^{\mp}$ & & $(p_{H^+}-p_h)_{\mu}Z_W^{1/2} \left[  g_{H^+h W} Z_{H^+H^-}^{1/2}Z_{hh}^{1/2} + g_{G^+hW}Z_{G^+H^-}^{1/2} Z_{hh}^{1/2} \right.$ \\ & & $ \left. + g_{G^+HW} Z_{G^+H^-}^{1/2}Z_{Hh}^{1/2}+g_{H^+HW} Z_{H^+H^-}^{1/2}Z_{Hh}^{1/2}+ \delta g_{H^+h W}  \right]$ \\ \\

$G^{\pm}h W_{\mu}^{\mp}$ & & $(p_{G^+}-p_h)_{\mu}Z_W^{1/2} \left[  g_{G^+h W} Z_{G^+G^-}^{1/2}Z_{hh}^{1/2} + g_{H^+hW}Z_{H^+G^-}^{1/2} Z_{hh}^{1/2} \right.$ \\ & & $ \left. + g_{G^+HW} Z_{G^+G^-}^{1/2}Z_{Hh}^{1/2}+g_{H^+HW} Z_{H^+G^-}^{1/2}Z_{Hh}^{1/2}+ \delta g_{G^+h W}  \right]$

\end{tabular}

\newpage
\begin{tabular}{lcl}
$G^{\pm}H W_{\mu}^{\mp}$ & & $(p_{G^+}-p_H)_{\mu}Z_W^{1/2} \left[  g_{G^+H W} Z_{G^+G^-}^{1/2}Z_{HH}^{1/2} + g_{H^+hW}Z_{H^+G^-}^{1/2} Z_{hH}^{1/2} \right.$ \\ & & $ \left. + g_{G^+hW} Z_{G^+G^-}^{1/2}Z_{hH}^{1/2}+g_{H^+HW} Z_{H^+G^-}^{1/2}Z_{HH}^{1/2}+ \delta g_{G^+H W}  \right]$ \\ \\

$H^{\pm}H W_{\mu}^{\mp}$ & & $(p_{H^+}-p_H)_{\mu}Z_W^{1/2} \left[  g_{H^+H W} Z_{H^+H^-}^{1/2}Z_{HH}^{1/2} + g_{G^+hW}Z_{G^+H^-}^{1/2} Z_{hH}^{1/2} \right.$ \\ & & $ \left. + g_{G^+HW} Z_{G^+H^-}^{1/2}Z_{HH}^{1/2}+g_{H^+hW} Z_{H^+H^-}^{1/2}Z_{hH}^{1/2}+ \delta g_{H^+H W}  \right]$
\end{tabular}

\subsubsection{1 scalar + 2 fermions}

In this section we present the counterterms for the scalar-fermion interactions. For the interactions with the neutral particles, $\psi_i$, will stand for up and down quarks, and charged leptons. For the interactions with the charged scalar particles we will use upper-case letters for fermions with $I_3=-1/2$ and lower-case for $I_3=1/2$. To simplify the form of the counterterms $[g_{ijk}]_L$ will stand for the left part of the coupling (proportional to $\gamma_L$) and  $[g_{ijk}]_R$ will stand for the right part of the same coupling. For the leptons, one of the couplings has to be set to zero by the reader. We also define the following quantities:

\begin{equation}
Z_{\psi}^{1/2}=Z_L^{1/2} \gamma_L + Z_R^{1/2} \gamma_R
\end{equation}
\begin{equation}
Z_{\overline{\psi}}^{1/2}=[Z_L^{\dagger}]^{1/2} \gamma_R + [Z_R^{\dagger}]^{1/2} \gamma_L \enskip .
\end{equation}

The counterterms are:

\begin{tabular}{lcl}
$\overline{\psi}_k \psi_l h$ & & $g_{\overline{\psi}_i \psi_i h}[Z_{\overline{\psi}}^{1/2}]_{ki} [Z_{\psi}^{1/2}]_{il}Z_{hh}^{1/2} + g_{\overline{\psi}_i \psi_i H} [Z_{\overline{\psi}}^{1/2}]_{ki} [Z_{\psi}^{1/2}]_{il}Z_{Hh}^{1/2} + \delta g_{\overline{\psi}_i \psi_i h}$
\end{tabular}

\newpage
\begin{tabular}{lcl}
$\overline{\psi}_k \psi_l H$ & & $g_{\overline{\psi}_i \psi_i H}[Z_{\overline{\psi}}^{1/2}]_{ki} [Z_{\psi}^{1/2}]_{il}Z_{HH}^{1/2} + g_{\overline{\psi}_i \psi_i h} [Z_{\overline{\psi}}^{1/2}]_{ki} [Z_{\psi}^{1/2}]_{il}Z_{hH}^{1/2} + \delta g_{\overline{\psi}_i \psi_i H}$ \\ \\

$\overline{\psi}_k \psi_l A$ & & $g_{\overline{\psi}_i \psi_i A}[Z_{\overline{\psi}}^{1/2}]_{ki} [Z_{\psi}^{1/2}]_{il}Z_{AA}^{1/2} + g_{\overline{\psi}_i \psi_i G_0} [Z_{\overline{\psi}}^{1/2}]_{ki} [Z_{\psi}^{1/2}]_{il}Z_{G_0 A}^{1/2} + \delta g_{\overline{\psi}_i \psi_i A}$ \\ \\

$\overline{\psi}_k \psi_l G_0$ & & $g_{\overline{\psi}_i \psi_i G_0} [Z_{\overline{\psi}}^{1/2}]_{ki} [Z_{\psi}^{1/2}]_{il}Z_{G_0 G_0}^{1/2} + g_{\overline{\psi}_i \psi_i A} [Z_{\overline{\psi}}^{1/2}]_{ki} [Z_{\psi}^{1/2}]_{il}Z_{AG_0}^{1/2} + \delta g_{\overline{\psi}_i \psi_i G_0}$ \\ \\

$\overline{\psi}_I \psi_i H^-$ & & $[Z_R^{\dagger}]^{1/2}_{IJ}[Z_L]^{1/2}_{ji}\left\{ [g_{\overline{\psi}_J \psi_j H^-}]_L Z_{H^+ H^-}^{1/2} + [g_{\overline{\psi}_J \psi_j G^-}]_L Z_{G^+H^-}^{1/2} \right\}$\\ \\ & & $+ [Z_L^{\dagger}]^{1/2}_{IJ}[Z_R]^{1/2}_{ji}\left\{ [g_{\overline{\psi}_J \psi_j H^-}]_R Z_{H^+ H^-}^{1/2} + [g_{\overline{\psi}_J \psi_j G^-}]_R Z_{G^+H^-}^{1/2} \right\} + \delta g_{\overline{\psi}_J \psi_j H^-}$ \\ \\

$\overline{\psi}_i \psi_I H^+$ & & $[Z_R^{\dagger}]^{1/2}_{ij}[Z_L]^{1/2}_{JI}\left\{ [g_{\overline{\psi}_j \psi_J H^+}]_L Z_{H^+ H^-}^{1/2} + [g_{\overline{\psi}_j \psi_J G^+}]_L Z_{G^+H^-}^{1/2} \right\}$\\ \\ & & $+ [Z_L^{\dagger}]^{1/2}_{ij}[Z_R]^{1/2}_{JI}\left\{ [g_{\overline{\psi}_j \psi_J H^+}]_R Z_{H^+ H^-}^{1/2} + [g_{\overline{\psi}_j \psi_J G^+}]_R Z_{G^+H^-}^{1/2} \right\} + \delta g_{\overline{\psi}_j \psi_J H^+}$ \\ \\

$\overline{\psi}_I \psi_i G^-$ & & $[Z_R^{\dagger}]^{1/2}_{IJ}[Z_L]^{1/2}_{ji}\left\{ [g_{\overline{\psi}_J \psi_j G^-}]_L Z_{G^+ G^-}^{1/2} + [g_{\overline{\psi}_J \psi_j H^-}]_L Z_{H^+G^-}^{1/2} \right\}$\\ \\ & & $+ [Z_L^{\dagger}]^{1/2}_{IJ}[Z_R]^{1/2}_{ji}\left\{ [g_{\overline{\psi}_J \psi_j G^-}]_R Z_{G^+ G^-}^{1/2} + [g_{\overline{\psi}_J \psi_j H^-}]_R Z_{H^+G^-}^{1/2} \right\} + \delta g_{\overline{\psi}_J \psi_j G^-}$ \\ \\

$\overline{\psi}_i \psi_I G^+$ & & $[Z_R^{\dagger}]^{1/2}_{ij}[Z_L]^{1/2}_{JI}\left\{ [g_{\overline{\psi}_j \psi_J G^+}]_L Z_{G^+ G^-}^{1/2} + [g_{\overline{\psi}_j \psi_J H^+}]_L Z_{H^+G^-}^{1/2} \right\}$\\ \\ & & $+ [Z_L^{\dagger}]^{1/2}_{ij}[Z_R]^{1/2}_{JI}\left\{ [g_{\overline{\psi}_j \psi_J G^+}]_R Z_{G^+ G^-}^{1/2} + [g_{\overline{\psi}_j \psi_J H^+}]_R Z_{H^+G^-}^{1/2} \right\} + \delta g_{\overline{\psi}_j \psi_J G^+}$ \\ \\
\end{tabular}

\newpage
\renewcommand{\thetable}{\Roman{table}}
\begin{table}[h]
\begin{center}
\begin{tabular}{|c|c|c|} \hline \hline
			     & \textbf{B\"ohm et al.} & \textbf{Aoki et al.} \\ \hline \hline
$\cal {L_{YM}}$ & $Z_W, Z_B, \delta e$ & $Z_W, Z_{ZZ}, Z_{ZA}, Z_{AZ}, Z_{AA}$ \\ & & $\delta M_W^2, \delta M_Z^2,\delta e$ \\ \hline
$\cal {L_{GF}}$ & $\delta \xi_i^W, \delta \xi_i^3, \delta \xi_i^B, \quad i=1, 2$ & 0 \\ \hline
$\cal {L_{S}}$     & $Z_{\phi}, \delta v, \delta \mu^2, \delta \lambda$ & 
$Z_H, Z_{G_0}, Z_{G^+}, \delta M_H^2, T$ \\ \hline
TOTAL                 & 13 & 13 \\ \hline \hline   
\end{tabular}
\caption{The Renormalization schemes of B\"ohm et al. and Aoki et al.}
\end{center}
\end{table}

\thispagestyle{empty}

\newpage

\begin{table}[h]
\begin{center}
\begin{tabular}{|c|c|c|c|c|} \hline \hline
			     & \textbf{Mod. I} & \textbf{Mod. II} & \textbf{Mod. III} & \textbf{Mod. IV} \\ \hline \hline
$\alpha_{eh}$ & $-\frac{\cos \alpha}{\sin \beta}$ & $\frac{\sin \alpha}{\cos \beta}$ & $-\frac{\cos \alpha}{\sin \beta}$ & $\frac{\sin \alpha}{\cos \beta}$ \\ \hline
$\alpha_{dh}$ & $-\frac{\cos \alpha}{\sin \beta}$ & $-\frac{\cos \alpha}{\sin \beta}$ & $\frac{\sin \alpha}{\cos \beta}$ & $\frac{\sin \alpha}{\cos \beta}$ \\ \hline
$\alpha_{eH}$ & $-\frac{\sin \alpha}{\sin \beta}$ & $-\frac{\cos \alpha}{\cos \beta}$ & $-\frac{\sin \alpha}{\sin \beta}$ & $-\frac{\cos \alpha}{\cos \beta}$ \\ \hline
$\alpha_{dH}$ & $-\frac{\sin \alpha}{\sin \beta}$ & $-\frac{\sin \alpha}{\sin \beta}$ & $-\frac{\cos \alpha}{\cos \beta}$ & $-\frac{\cos \alpha}{\cos \beta}$ \\ \hline
$\beta_{e}$ & $-\cot \beta$ & $\tan \beta$ & $-\cot \beta$ & $\tan \beta$ \\ \hline
$\beta_{d}$ & $-\cot \beta$ & $-\cot \beta$ & $\tan \beta$ & $\tan \beta$ \\ \hline
\end{tabular}
\caption{Coupling constants for the fermion-scalar interactions}
\end{center}
\end{table}

\thispagestyle{empty}

\newpage

\textbf{FIGURE CAPTIONS}

Figure 1. WG and WH mixing.

Figure 2. The tadpole condition. 

\thispagestyle{empty}

\newpage

\bigskip
\bigskip

\begin{figure}[h]
\begin{center}
\setlength{\unitlength}{1.5mm}
\thicklines
$W^+$ \begin{picture}(30,1)(0,0)
\multiput(0,0)(3,0){3}{\begin{picture}(3,1)(0,0)
\put(0,0){\line(1,0){0.75}}
\put(2.25,0){\line(1,0){0.75}}
\end{picture}}
\put(15,0){\circle{12}}
\multiput(21,0)(3,0){3}{\begin{picture}(3,1)(0,0)
\put(0,0){\line(1,0){0.75}}\put(2.25,0){\line(1,0){0.75}}
\end{picture}}
\end{picture}
\enskip $G^+ ,\enskip H^+$
\end{center}
\end{figure}

\thispagestyle{empty}

\newpage

.

\bigskip
\bigskip
\bigskip
\bigskip
\bigskip
\bigskip
\bigskip
\bigskip
\bigskip
\bigskip

\begin{figure}[h]
\begin{center}
\setlength{\unitlength}{1.5mm}
\thicklines
\begin{picture}(10,1)(0,0)
\multiput(0,0)(0,3){4}{\begin{picture}(3,1)(0,0)
\put(0,0){\line(0,1){0.75}}
\end{picture}}
\put(0,15){\circle{12}}
\put(12,10){+}
\put(35,10){= \quad 0}
\put(4,4){H,\enskip h}
\put(25,4){H,\enskip h}
\end{picture}
\begin{picture}(10,1)(-10,0)
\multiput(0,0)(0,3){5}{\begin{picture}(3,1)(0,0)
\put(0,0){\line(0,1){0.75}}
\end{picture}}
\put(-1.8,12){\huge X}  
\end{picture}
\end{center}
\end{figure} 

\thispagestyle{empty}


\newpage
\begin{thebibliography}{99}

\bibitem{SMP}J. Busenitz, Proc. of the Inter. Europhysics Conf. on High \\
			Energy Physics, Brussels 1995, J. Lemonne, C. V. Velde and \\
			F. Verbeure Edts. , World Scientific (1995) 37; \\
			D. Bardin, \textit{ibid.} pg. 41.
\bibitem{Ren}G.'t Hooft, Nucl. Phys. \textbf{B33} (1971) 133; \\
			 B. W. Lee and J. Zinn-Justin, Phys. Rev. \textbf{D5}, (1972) 3121; \\
			 D. A. Ross and J. C. Taylor, Nucl. Phys. \textbf{B51}  (1973) 125. 
\bibitem {Aoki}K. Aoki, Z. Hioki, R. Kawabe, M. Konuma, T. Muta, \\
			Suppl. Prog. Theor. Phys. \textbf{73} (1982) 1.
\bibitem {GHDK90} J. Gunion, H. Haber, G. Kane and S. Dawson, \\
				 The Higgs Hunter's Guide, Addison Wesley (1990).
\bibitem {CDF} F. Abe \textit{et al} [CDF collaboration], Phys. Rev. Lett. \textbf{74} (1995) 2626; \\
			 S. Abachi \textit{et al} [D0 collaboration], Phys. Rev. Lett. \textbf{74} (1995) 2632.
\bibitem {MP92}A. M\'endez and A. Pomarol, Phys. Lett. \textbf{B279} (1992) 98. 
\bibitem{RR92}S. Raychaudhuri and A. Raychaudhuri, Phys. Lett. \textbf{B297} (1992) 159; \\
			 J. Gunion, G. Kane and J. Wudka, Nucl. Phys. \textbf{B299} (1988) 231; \\
			 M. Capdequi-Peyran\^ere, H. Haber and P. Irulegui, Phys. Rev. \textbf{D44} (1991) 191.
\bibitem {PP93} D. Pierce and A. Papadopoulos, Phys. Rev. \textbf{D47} (1993) 222.
\bibitem{CPR94}P. H. Chankowski, S. Pokorski and J. Rosiek, Nucl. Phys. \textbf{B423} (1994) 437.
\bibitem{L73}T. D. Lee. Phys. Rev. \textbf{D8} (1973) 1226.
\bibitem{rui1} J. Velhinho, R. Santos and A. Barroso, Phys. Lett. \textbf{B322} (1994) 213. 
\bibitem{Hollik}W. Hollik, Proc. of the Inter. Europhysics Conf. on High Energy \\
			Physics, Brussels 1995, J. Lemonne, C. V. Velde and  F. Verbeure \\
			Edts., World Scientific (1995) 921,\textit{and references therein}
\bibitem{BRS74} C. Bechi, A. Rouet and R. Stora, Phys. Lett., \textbf{52B} (1974) 344.\\
			    I. V. Tyutin, Lebedev Report No. FIAN 39, 1975 (in Russian) (unpublished).
\bibitem{B85} L. Baulieu, Phys. Rep. \textbf{129} (1985) 1.
\bibitem{FP67}  L. D. Faddeev and V. N. Popov, Phys. Lett. \textbf{25B} (1967) 29.
\bibitem{BHS87} M. B\"ohm, W. Hollik and H. Spiesberger, Fortsch. Physt. \textbf{34} (1987) 687.
\bibitem{RT73} D. A. Ross and J. C. Taylor, Nucl. Phys. \textbf{B51} (1973) 125.
\bibitem{SN94}A. Schilling and P. van Nieuwenhuizen, Phys. Rev. \textbf{D50} (1994) 967.
\bibitem{DS90} A. Denner and T. Sack, Nucl. Phys. \textbf{B347} (1990) 204.
\bibitem{rui2} R. Santos, A. Barroso and L. Br\"ucher, Univ. Mainz \\ 
		       preprint MZ-TH 96-16, (to be published in Phys. Lett. B).
\end{thebibliography}
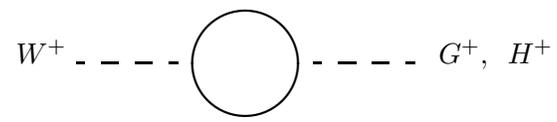
\end{document}